\documentclass{article}

\usepackage[preprint,nonatbib]{neurips_2020_changed}

\usepackage[utf8]{inputenc} % allow utf-8 input
\usepackage[T1]{fontenc}    % use 8-bit T1 fonts
\usepackage{hyperref}       % hyperlinks
\usepackage{url}            % simple URL typesetting
\usepackage{booktabs}       % professional-quality tables
\usepackage{amsfonts}       % blackboard math symbols
\usepackage{nicefrac}       % compact symbols for 1/2, etc.
\usepackage{caption}
\usepackage{microtype}      % microtypography
\usepackage{soul} % for \hl
\usepackage{color} 
\usepackage{paralist}  % compact itemize
\usepackage{graphicx}
\usepackage{amsmath}
\usepackage{amsfonts} % For notation R - real number space
\usepackage{colortbl} % for colored cells in tables
\usepackage{graphicx} % For Table 1 - multiple row titles
\usepackage{multirow} % For Table 1 - multiple row titles

\title{Deep Negative Volume Segmentation}

\def\inst#1{\unskip$^{#1}$}
\author{%
    Kristina Belikova\inst{1} \\
    \texttt{kristina.belikova@skoltech.ru} \\
    \And
    Oleg Rogov\inst{1} \\
    \texttt{o.rogov@skoltech.ru} \\
    \AND
    Aleksandr Rybakov\inst{2} \\
    \texttt{rybakov.aleksandr@gmail.com} \\
    \And
    Maxim V. Maslov\inst{2} \\
    \texttt{maximmaslov@mail.ru} \\
    \AND
    Dmitry V. Dylov\inst{1} \\
    \texttt{d.dylov@skoltech.ru} \\
    \\
  \unskip$^{1}$ Skolkovo Institute of Science and Technology\\
  \unskip$^{2}$ Pavlov First St.Petersburg State Medical University \\
}

\begin{document}

\maketitle

\begin{abstract}
Clinical examination of three-dimensional image data of compound anatomical objects, such as complex joints, remains a tedious process, demanding the time and the expertise of physicians. For instance, automation of the segmentation task of the TMJ (temporomandibular joint) has been hindered by its compound three-dimensional shape, multiple overlaid textures, an abundance of surrounding irregularities in the skull, and a virtually omnidirectional range of the jaw's motion -- all of which extend the manual annotation process to more than an hour per patient. To address the challenge, we invent a new angle to the 3D segmentation task: namely, we propose to segment empty spaces between all the tissues surrounding the object -- the so-called negative volume segmentation. Our approach is an end-to-end pipeline that comprises a V-Net for bone segmentation, a 3D volume construction by inflation of the reconstructed bone head in all directions along the normal vector to its mesh faces. Eventually confined within the skull bones, the inflated surface occupies the entire ``negative'' space in the joint, effectively providing a geometrical/topological metric of the joint's health. We validate the idea on the CT scans in a 50-patient dataset, annotated by experts in maxillofacial medicine, quantitatively compare the asymmetry given the left and the right negative volumes, and automate the entire framework for clinical adoption.
\end{abstract}

\section{Introduction}

Our study began from the following simple question while we were performing a very tedious manual annotation of a compound three-dimensional (3D) structure. \textbf{Q}: Instead of finding the exact contours that circumscribe the 3D object, can we segment \emph{the air} that fills the gaps within its parts? What deep neural network architecture would accomplish that, given the gaps are the \emph{absolute complements} to the annotation labels? To find answers, we geared up with the most complex 3D object we could find.

% Despite the excellent quality of the data captured by the modern imaging modalities (e.g., X-ray, CT, MRI), the evaluation of the images of complex joints is still primarily a manual task.
%  
Some of the most structurally complex objects in the human body are indisputably the joints, in general, and the \textit{temporomandibular joint\footnote{TMJ is a bilateral joint formed by the \emph{mandibular} and the \emph{temporal} bones of the skull, differing from the other joints anatomically and functionally~\cite{TMJ_mechanics,Wadhwa930_TMJ_disorders}. 
TMJs enable functions like chewing and speaking.}} (TMJ), in particular. Several medical research groups still actively debate trying to explain the kinetic function of the TMJ joint, its multiple degrees of freedom, and even its relation to a plethora of known illnesses (maxillofacial ones and beyond~\cite{Wadhwa930_TMJ_disorders,TMJ_disorder_qualityLife}). 
Accurate interpretation of TMJ images has become essential in a variety of clinical practices, ranging from the basic assessment of wear and tear (e.g., osteoarthritis) to intricate surgical interventions (e.g., arthroplasty). 
The lack of trustworthy automation of the basic diagnosis-assisting routines (such as tendon segmentation or a measurement of the cartilage wear) stems from the fact that such compound joints have extremely intricate 3D anatomy and a variety of surrounding tissues of perplexed morphologies and textures~\cite{Farias2015_morphology_var}. We show a number of 3D examples of the TMJ's complex geometries in the supplementary material.

Millions of people suffer from temporomandibular disorders (TMDs), having such symptoms as a limitation or a deviation of the range of the jaw's motion, certain TMJ sounds, associated headache, and the very pain in the joints. Orthodontic, maxillofacial, and plastic surgeries cover the other large related cohort of patients. Despite being that common, the diagnostics of all of the mentioned TMJ symptoms remains very challenging~\cite{Talmaceanu2018_disorders_tmj}, and the current clinical practice entails very rudimentary linear or 2D measurements of the joint's tissues.
% A TMJ is characterized by a complex anatomic structure and specific irregularities with the dimensions of the computed tomography (CT) spatial resolution limits, and is among the most complex joints in human body with vast morphological variability \cite{Farias2015_morphology_var}. 
% Because of the complexity of TMJ, the use of 2D slice-by-slice visualization is not sufficient, requiring a true 3D reconstruction to describe its anatomy and to find the cause of a given symptom. Yet, many dentists, especially in the developing countries, have to dismiss the 3D structure and to resort to simple linear measurements of the object dimensions in the 2D images.
Such measurements have obvious shortcomings: they are subjective, time-consuming, and not accurate enough due to the in-plain estimations.
In fact, significant outcome differences were reported when TMJ is measured in 2D \textit{vs.} in 3D~\cite{2d_vs_3d}. True 3D characterization of TMJ in medical images is essential for improving various clinical practices, including dentistry, orthodontics, maxillofacial and plastic surgeries. 

Manual 3D annotation of the TMJ is usually undertaken only by the top hospitals, requiring expertise of the maxillofacial doctors, that of a 3D modelling technician, and a long collaborative effort to draw a fitting 3D model of the jaw and of the other head parts involved \cite{Nove3DTMJ}. In fact, there is simply \emph{no standardized annotation workflow} for contouring the TMJ structures \emph{even manually} today. In this work, we propose such a workflow, by suggesting to shift the focus from the segmentation of the hard-to-contour anatomical structures within the joint to the segmentation of the spaces between these structures (the gaps). We have called the method ``negative volume'' segmentation and presented a new method of manually annotating such a volume in Section \ref{sec:manual}. Also, we present an end-to-end pipeline for deep negative volume segmentation to automate and to improve the manual one. The fully-automatic 3D deep negative volume segmentation approach is described in Section~\ref{sec:automatic_pipeline}.

\textbf{Contributions.} The key contributions of our paper are the following:
\begin{compactitem}
\item New paradigm for segmentation of the `air gaps' within complex 3D objects (the concept of ``Negative Volume'') using a deep neural network.
\item New \emph{manual} annotation workflow for negative volume segmentation in the human joints. It is multiple orders of magnitude more descriptive than current clinical standard.
\item First \emph{automatic} end-to-end pipeline for extraction of negative volumes within a human's joint, incorporating deep learning-based localization, segmentation, and surface mesh inflation.
\item New volumetric measure of a joint's health based on its symmetry properties via the state-of-the-art topological cloud-to-cloud metrics.
\end{compactitem}
% consuming task and requires long training and high skills due to the necessity of annotating skull bones, articular disk, and the mandible bone prior to the segmentation of the latter. 
% Due to the absence of proper expert annotation and shape complexity, TMJ segmentation remains a challenging problem and improving the automatic segmentation of TMJ from CT scans is particularly significant for the dental community. 

\section{Related work}
\textbf{Joint space assessment} 
 is essential in many clinical practices (TMJs, wrist joints, shoulders, knee joints, etc.), ranging from orthopedics to plastic surgeries~\cite{orthopedics,plastic_surgery}. While different metrics of inter-articular space are calculated for some joints, the exact definition of the joint space boundaries is still a matter of debate. 
We refer to ~\cite{wrist_review} (generic joint papers), to ~\cite{tmj_space} (TMJ review papers), and to our supplementary material (a comprehensive review of relevant medical literature).
% and the state-of-the-art in clinical practice).

\textbf{Object localization on medical scans.} 
Automatic localization of objects of interest is a prerequisite for many medical imaging tasks, as it can narrow down the field of view to the important structures. As of today, there are several approaches for detecting specific areas of various shapes and sizes such as body parts, bone tissues, organs, nodules, and tumors in 3D MRI and CT images~\cite{SHEN2017663, cervical_vertebrae, roi_aware_unet, prokopenko2019unpaired, skull_segmentation}. Completely autonomous cropping in medical images has been reported~\cite{SHEN2017663}. It is a common practice to use a cascaded approach, consisted of several steps: object localization and object segmentation or another required action. The first step is to localize the object from the entire 3D scan, and then provide a reliable bounding box for the more refined steps~\cite{multi_task_cascades}, Mask R-CNN~\cite{mask_rcnn}, 3D RoI-aware U-Net~\cite{roi_aware_unet}, segmentation-by-detection~\cite{segmentation_by_detection}, etc.).

% \hl{Architectures based on the Convolutional Neural Networks and related to localization of compound joints i.e. such as knee}~\cite{knee_segmentation, knee_box}
% \hl{and hips}~\cite{segmentation_by_detection}
% \hl{are of special interest for our task, since the aim is to localize junction between bones (small adjoining parts of the bones) and not the bone structures entirely as in case of organs or tumors. These approaches aimed to automatically detect joint region utilize a coarse segmentation to solve the localization problem.}

\textbf{Medical image segmentation.} With the advent of artificial intelligence to medical image computing, a wide range of image segmentation challenges were successfully tackled by deep learning methods~(see Refs.~\cite{Lee-DL-overview,chowdhury2017blood,Zhou2019} for review). In particular, significant advances were made by the architectures based on the Convolutional Neural Networks (U-Net \cite{ronneberger2015unet,iek2016_3dUnet}, 
V-Net \cite{milletari2016_Vnet}, 
U-Net++ \cite{zhou2018unetPlusPlus}, 
MD U-Net \cite{zhang2018mdunet}, 
Stack U-Net \cite{Sevastopolsky2019}, etc.).
Of specific value to our task, are the 3D U-Net~\cite{iek2016_3dUnet} and the attention-gated 3D U-Net~\cite{oktay2018_attentionUnet} architectures that take advantage of efficient GPU computing, the ability to achieve high precision with a fewer training samples, and the capability of ``learning where to look'' with the class-specific pooling \cite{MulticlassWTMJ}. To automate the negative volume segmentation task, we first needed to segment the major bones (mandibular and temporal bones), which eventually draw us to select the V-Net architecture~\cite{milletari2016_Vnet}. V-Net is similar to 3D U-Net but is more prone to convergence thanks to learning the residual function along the way. The summary of the architecture selection is covered in Section~\ref{sec:results}.

Once the bone segmentation was automated, we proceeded with the segmentation of the space between the bones. For that, we introduced a new \textit{inflation} procedure that gradually fills the space between the inner structures of the joint until the entire negative volume is occupied.  The inflation procedure and the full segmentation pipeline are described in Section \ref{sec:inflate}.
% Resolution is reduced by convolution with 2x2x2 kernels applied with stride of 2 to have a smaller memory footprint during training. Deconvolution is employed in order to increase the size of sample \cite{Milletari2016_VNET}. 

\textbf{Mesh inflation} 
% change first citation
Deformation, inflation or deflation are commonly employed in complex 3D reconstruction problems to boost the model quality by detailing the meshes. Modern physics-based mesh deformation and generation methods, combine robust constraint optimization and efficient re-meshing \cite{ACCVinflation}, which proved useful in medical imaging~\cite{CountourSegment2018} but still requires additional evaluation of the nesting feasibility criteria, often viewed as constraint optimization problems for meshes~\cite{NestedMesh2017}. 
\section{Methods}
\begin{figure}[b]
    \centering
     \includegraphics[width=0.95\textwidth]{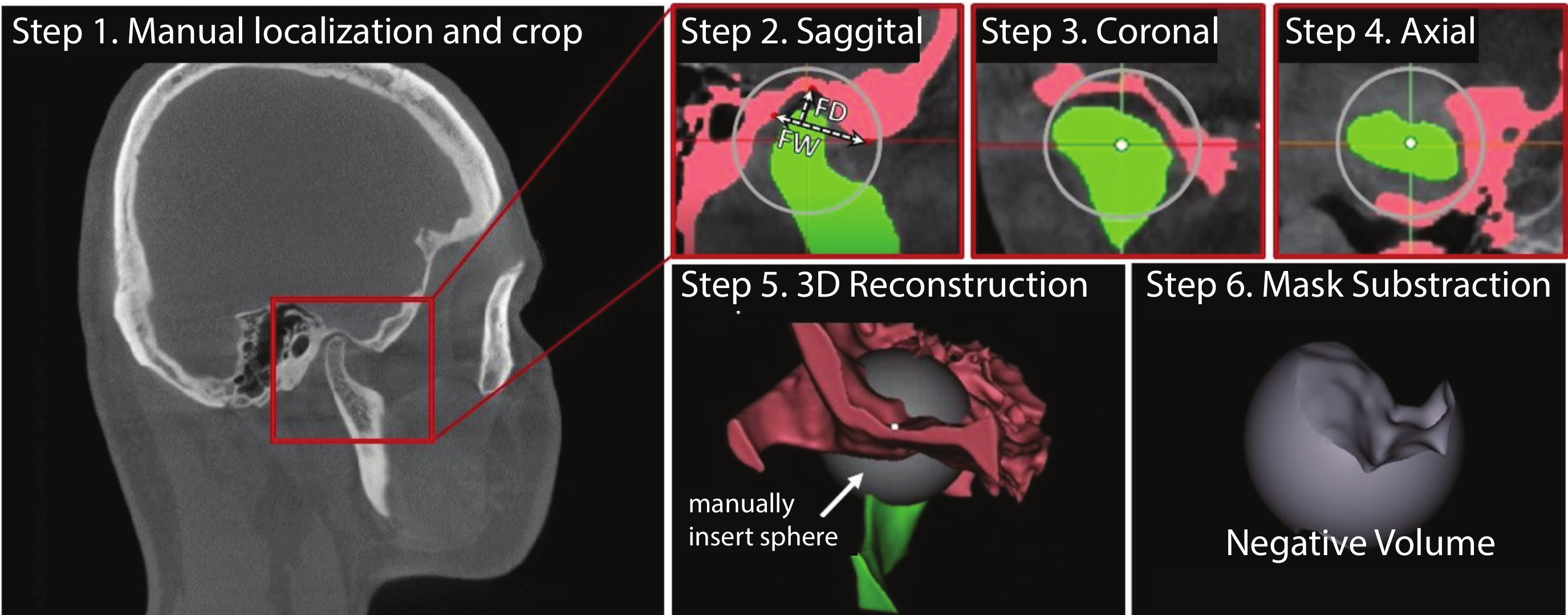}
    \caption{Proposed steps for manual negative volume annotation in TMJ (left  to right). The process requires drawing masks around complex structures of mandibular condyle (green) and temporal bones (red) in all three views (saggital, coronal, and axial) for each slice of the volume of interest (VOI), until the resulting 3D reconstruction allows to subtract the negative ``ball'' from a manually inserted sphere. Such annotation \emph{takes about 1 hour per patient}.}
    \label{fig:pipeline_annot}
\end{figure}
% \vspace{-0.1cm}
This Section covers the concept and the workflow to generate negative volumes via two pipelines: manual 3D annotation (Section~\ref{sec:manual}) and an end-to-end automatic approach which is even more descriptive than the proposed manual one (Section~\ref{sec:automatic_pipeline}), suggesting a new metric for the joints.
% In the final part of the Section, we emphasize the fundamental difference between manual annotation and automatic approach -- manual annotation, like all previous methods of joint space evaluation, does not have sufficient accuracy in determining the clear boundaries of the joint space. 
\subsection{Manual Annotation Pipeline: Negative Volume Concept}
\label{sec:manual}
To reveal the concept of negative volume, we introduce a new method for examination of complex joints that takes advantage of all available 3D information acquired by an imaging modality. Fig.~\ref{fig:pipeline_annot} proposes \emph{volumetric characterization} of a joint, with TMJ taken as an example. The method targets extraction of the \emph{empty space} between the various tissues surrounding the joint, which we intuitively call a ``negative volume''.
To extract it, the proposed manual annotation pipeline entails drawing a series of 2D masks for the mandibular condyle (MC) and for the temporal bone (TB) in a cropped sequence of the original DICOM, a resulting 3D reconstruction of the volumes of the MC and TB bones, a manual (rough) positioning of a 3D sphere within the joint center, and a consequent subtraction of the mask volumes from the sphere.

Unlike the current clinical examinations~\cite{TMJ_FW_lateral}, where the width and the depth of the mandibular fossa are measured (``FW'' and ``FD'' in Fig.~\ref{fig:pipeline_annot}), the true volumetric ``negative ball'' extracted from the joint is far more informative. It takes more than an hour to annotate one patient; if automated, it could be quickly adopted in the clinical practice as a new measure of joint's health.

% We use a V-Net architecture for bone structure segmentation.
%
%
%
%
%
%
\subsection{Automatic Pipeline: Segmentation of Negative Volume}
\label{sec:automatic_pipeline}
\begin{figure}[b]
    \centering
     \includegraphics[width=0.99\textwidth]{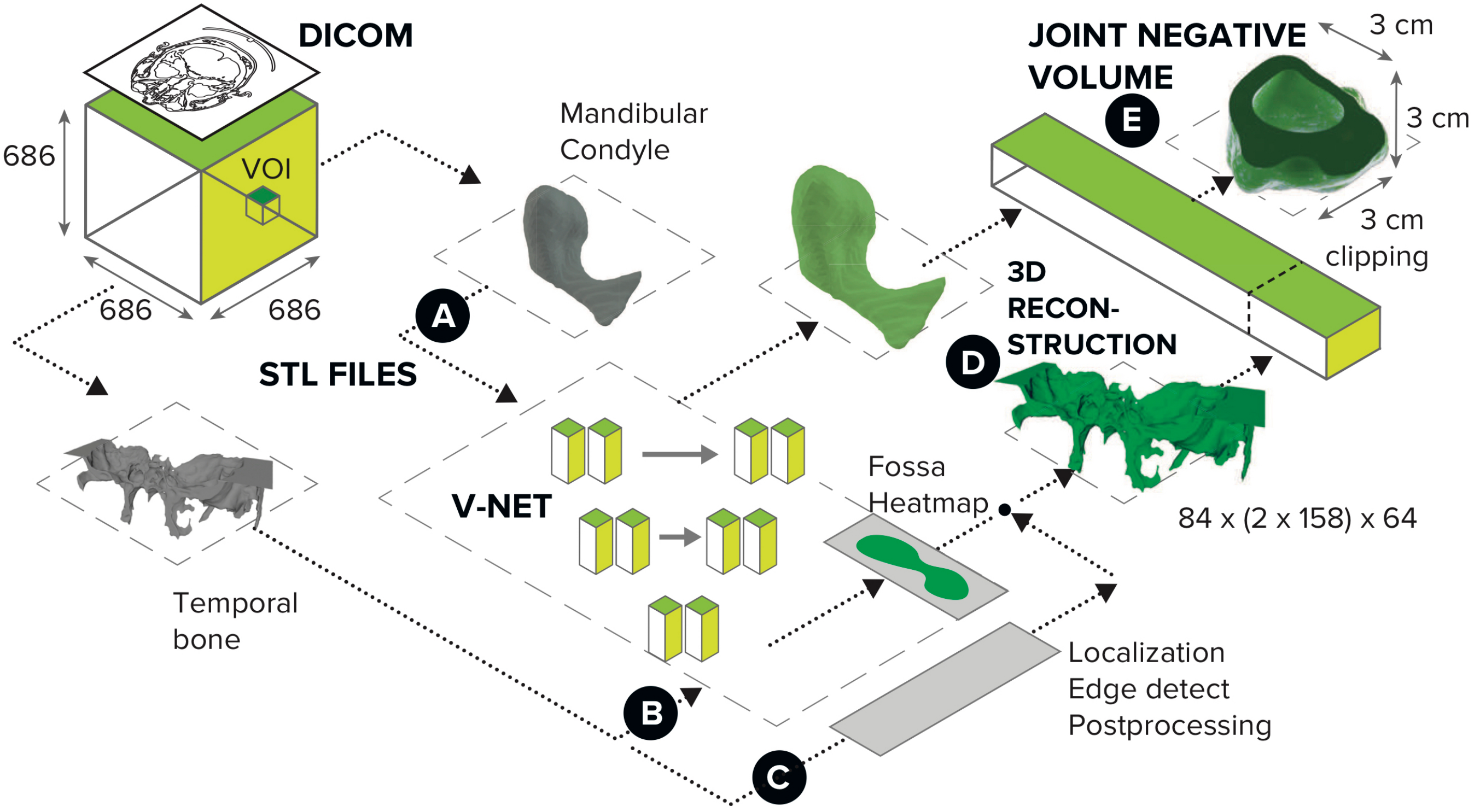}
    \caption{End-to-end pipeline for Deep Negative Volume Segmentation in joints. Segmentation of MC and TB are shown as step A and step B, respectively. Step C and step D represent classical image enhancement of both bone reconstructions. Fig.~\ref{fig:inflation} shows ``inflation/clipping'' block (step~E) in detail.}
    \label{fig:pipelie_Vnet}
\end{figure}

We now proceed to automating an end-to-end pipeline based on the approach in Fig.~\ref{fig:pipeline_annot} but with several principle differences which stem from the fact that such negative volumes are impossible to annotate in a sufficient number manually (to train a basic 3D network).
%for the implementation of the approach in Fig.~\ref{fig:pipeline_annot}. 
The proposed pipeline consists of the following steps: data preprocessing, volume of interest (VOI) selection, segmentation of the TB and MC bones, 3D reconstruction of the segmentation results, inflation of the MC volume to fit into the mandibular fossa, and, finally, extraction of the negative volume by clipping (see Fig.~\ref{fig:pipelie_Vnet}).

% \vspace{-0.5cm}
% Сlassical image enhancement of TB (step C) is performed in parallel with step B and along with step A is followed by 3D Reconstruction on both bones (step D)
% \vspace{-1cm}
\paragraph{Data preprocessing} Basic DICOM data normalization and confirmation of the co-alignment of the ground truth annotation masks are done as the first step. 
% Because the CT scans were acquired on a Planmeca ProMax scanner with the same contrast settings and spatial resolution, no special normalization in the Hounsfield units (HU) was required in our case. 
The data preprocessing consisted of min-max normalization of DICOM data and voxelization of Standard Triangle Language (STL) models.
Details of STL models voxelization and further data augmentation are given in Section~\ref{sec:experiments}
%Details of STL mask pre-processing are given in Section~\ref{sec:dataset}.

% \vspace{-0.25cm}
\paragraph{VOI selection}
% \hl{Before proceeding with segmentation of the main bones (mandibular and temporal bones) inside the joint, we first need to localize the VOI from the 3D scans, which should completely contain the area of the joint. This step is justified by the fact that it is hardly possible to train a voxel-based model with high-resolution inputs e.g., V-Net despite a relatively small memory footprint requires about 1156 GB of GPU memory on our $686\times686\times686$ inputs with batch size of 1 and 3D U-Net requires 1286 GB under the same conditions.}

We have approached the localization of TMJ VOI bounding the bones (MC and TB) as a segmentation problem at a lower resolution, based on the available memory and size of input data. To perform localization of joint we utilize V-Net model, which has proven itself as an accurate enough voxel-based model with fast convergence. For our case, we resize the raw images to a lower resolution $160\times 160 \times 160$ using bicubic interpolation to preserve available memory.
%A VOI bounding the bones (MC and TB) was trained by an auto-crop network combined with individual post-tuning of each scan to locate the anatomical reference points in the vicinity of the TMJ. We omit details of this obvious step and note that the post-tuning is the only part of our pipeline which we will eliminate later. Completely autonomous cropping in medical images has been reported~\cite{SHEN2017663} and should work with our data seamlessly once a larger dataset is available. 
%This step results in two cropped volumes to be used for training the main neural network: MC ($182 \times 154 \times 64$) and TB ($88 \times 136 \times 96$).
This step results in two cropped volumes of various sizes to be used for training the segmentation neural network: both the left and the right joints with separate masks for MC and TB.

% in $x \times y \times z$
% This means that only MC - small parts of the mandible were labeled, despite the fact that they do not have a clear distinguishable border with the rest of the mandible. 
% Similarly, the annotation for TB was incomplete. For this reason, choosing a single region for all CT results in poor quality of model training due to the inconsistent data examples. So we crop all CTs individually but in the same size to perform segmentation of MC and TB separately.
% The overall VOI includes the union of all these individual volumes including ones symmetrically reflected relative to the sagittal plane. In order to have additional spatial information about TMJ location, the final VOI was expanded by 15 voxels in all 6 directions. 
% Therefore, the result MC data sizes used in experiments are  
% and TB data sizes are $88 \times 136 \times 96$ in $x \times y \times z$ dimensions. 
%
% \vspace{-0.25cm}
\paragraph{3D Bone segmentation: (A) MC and (B) TB bones}\label{bone_segment}
% \subsubsection{3D Bone segmentation: TB and MC bones}
% \label{sec:segmselect} 
One has to resort to architectures for 3D segmentation due to the complex structure and texture of the bones in that part of the skull (especially, the TB which has many irregularities). The V-Net architecture proved to work best for the MC, 
%and the 3D U-Net with attention gate worked best 
as well as for the complex TB bone. Full comparison of the architectures is given in Table~\ref{tab:table_segmentation}, with V-Net being better for deployment due to its faster convergence (to segment both MC and TB).
% \vspace{-0.5cm}
\paragraph{(C) Classical image enhancement}
While MC segmentation via V-Net proved satisfactory (step A in Fig.~\ref{fig:pipelie_Vnet}), the TB segmentation (step B in Fig.~\ref{fig:pipelie_Vnet}) needed to be enhanced by passing the original data through a classical processing route (step C in Fig.~\ref{fig:pipelie_Vnet}): namely, we applied the removal of noise, closing edges, morphological smoothing (such as erosion and dilation), and 3D Canny edge detection filters. The sequence of these operations is completely automated and the result is fused with the \textit{fossa} heatmap, generated by V-Net, to provide a single TB mask. Notably, the step C could be removed altogether once a sufficiently large number of manual annotations of the TB is collected. 
Or, it can be viewed as 
% a data  augmentation to ``compensate'' 
''compensation'' for the complex irregularities encountered in the joint, which would otherwise require a lot of annotation for training. 
% . In case of temporal bone segmentation, we use classical edge detection approach to enhance the quality of bones segmentation (path B). For this purpose, we use outputs of the model as probability map of TB location and perform 3D Canny edge detection on the pixel-wise product of VOI and segmented output.  
% \vspace{-0.5cm}
\paragraph{(D) 3D reconstruction}
% TMJ cross-sections are acquired using CT tomograph (e.g. Philips or smth). 
To reconstruct 3D models, 116 equidistant consecutive sections with a pitch of 0.4 mm and a bounding box dimension of 103 px $\times$ 158 px were used.  The fused surfaces of the interfaces between the articular disk, MC, and TB were subjected to median averaging,
2D-filtering, and interpolation, entailing filter radius matching (to fit the size of irregularities) and the edge detection applied in a slice-by-slice manner.
% The radius of the filter is chosen accordingly to the size of the irregularities of interest as filtering reduces noise yet also may suppress small objects, which is crucial for the TMJ segmentation task.
% Another purpose of filtering is edge detection, in 2D a classical way to detect edges is to compute edges in both X and Y directions and then combine them to find the norm of the gradient corresponding to the strength of the edges, the angle of the edges can also be computed from the X and Y differentials. For 3D edges the differential in Z is added to the X and Y ones.
% \vspace{-0.25cm}
\begin{figure}[b]
    \centering
    % \includegraphics[width=0.85\textwidth]{EPS/INFLATE_VECTOR_thinner.pdf}
    % NORMAL BELOW ->>>>>>>>
    %  \includegraphics[width=0.95\textwidth]{EPS/INFLATE_VECTOR.pdf}
    \includegraphics[width=0.95\textwidth]{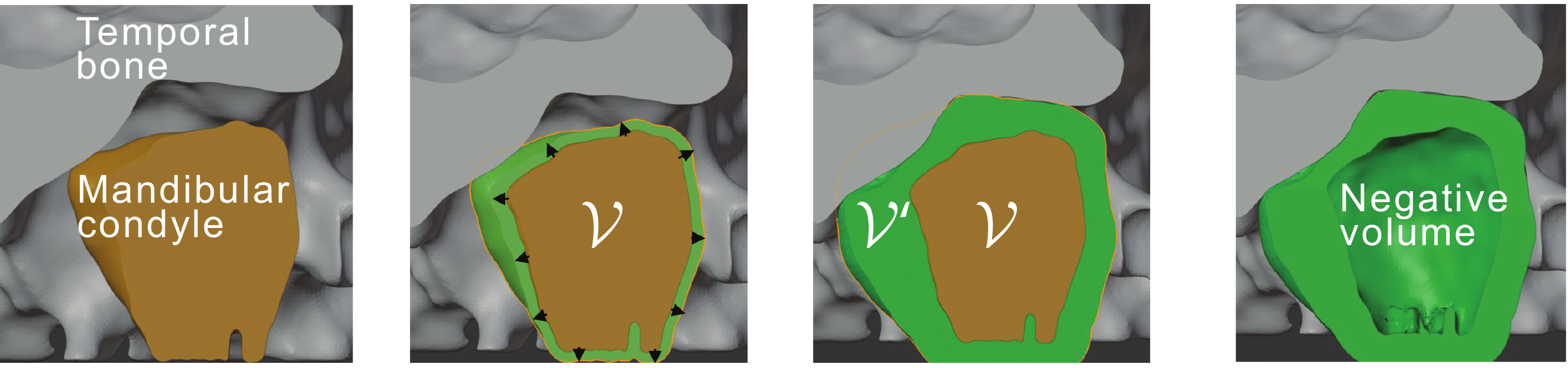}
    \caption{Proposed negative volume \emph{inflation routine} seen in TMJ cross-section (frontal view): (1) segmented MC bone is a starting point (mesh $\mathcal{V}$ ), (2) surface of MC spreads along the normals, (3) inflated MC reaches bounding volume defined by TB model (mesh $\mathcal{V}'$), (4) MC removal and clipping of the neck of the mandible generates the negative volume.}
    \label{fig:inflation}
\end{figure}
% \vspace{-1cm}
\paragraph{(E) Negative volume inflation}
\label{sec:inflate}
% Given a segmented volume, we obtain its triangular mesh $V$ in 3D space by means of the 3D reconstruction. 
Fig.~\ref{fig:inflation} summarizes how the mesh $\mathcal{V}$ of the reconstructed MC bone is inflated along the normals to maximize similarity with the \emph{fossa} space surrounding the TB mesh $\mathcal{V}'$. Inflating the mesh $\mathcal{V}$ belongs to a class of optimization problems that are accompanied by the Laplacian regularization to ensure a smoother shape \cite{Skouras2014_Inflation}. Boolean difference of the two meshes $\mathcal{V}'\setminus \mathcal{V}$ provides the final negative volume of interest.
% \vspace{-0.5cm}
\subsubsection{Symmetry metrics}
Having received both the left ($\mathcal{L}$) and the right ($\mathcal{R}$) negative volumes, the doctors can proceed to any accurate volumetric measurements, relevant to a given set of particular symptoms and conditions at hand. In maxillofacial practice, for instance, it is quite common to estimate the $\mathcal{L}$-$\mathcal{R}$ symmetry\cite{Talmaceanu2018_disorders_tmj} of the TMJs, which directly correlates with the jaw's alignment. For that, we suggest to use a volumetric measure based on the Hausdorff cloud-to-cloud distance. To estimate the symmetry between the two negative volumes, we define the Hausdorff distance for two point sets ($\mathcal{L}$ and $\mathcal{R}$) on a  metric space $(\mathbb{R}^3,d)$, where $d(l,r)$ is the Euclidean distance between the points $l$ and $r$.
% \begin{equation*}
%     H_{LR}=\max \left\{\sup _{{l} \in \mathcal{L}} \inf _{r \in \mathcal{R}} d(l, r), \sup _{r \in \mathcal{R}} \inf _{l \in \mathcal{L}} d(l, r)\right\}, \qquad 
%     S_{LR}= \frac{\max \left\{S_L, S_R\right\}}{\min \left\{S_L, S_R\right\}} 
% \end{equation*}
% where $d(l,r) = \sqrt{(x_l-x_r)^2+(y_l-y_r)^2+(z_l-z_r)^2}$ is the Euclidean distance between points $l$ and $r$. 
The Hausdorff measure is a well-known and a robust metric that exists in many programming libraries. Many other possible metrics could be also proposed though, e.g., $S_{LR}$ the ratio of the mesh surface areas of both
% the larger and of the smaller 
negative volumes 
%(S$_{LR}$), which we also report in the Results Section.
$S_L$ and $S_R$, 
where the lower index corresponds to the left and right volume respectively.
\begin{equation*}
    H_{LR}=\max \left\{\sup _{{l} \in \mathcal{L}} \inf _{r \in \mathcal{R}} d(l, r), \sup _{r \in \mathcal{R}} \inf _{l \in \mathcal{L}} d(l, r)\right\}, \qquad 
    S_{LR}= \frac{\max \left\{S_L, S_R\right\}}{\min \left\{S_L, S_R\right\}} 
\end{equation*}
% \begin{equation}
%     S_{LR}= \frac{\max \left\{S_L, S_R\right\}}{\min \left\{S_L, S_R\right\}},
% \end{equation}
We report measurements with both proposed symmetry metrics in the Results Section.
These metrics are as descriptive as possible and ought to replace the simplistic conventional linear measurements.
% \begin{equation}
    
% \end{equation}

% \begin{equation}
%     \bar{V} =\{p+\epsilon h(p) U | p \in V\}
% \end{equation}

% \subsubsection{Closing the surface of the bone}
% We use Laplacian smoothing over bone mesh in order to obtain proper surface roughness of the reconstructed volume [Hansen, Glen A.; Douglass]. On meshes smoothing can be resolved a as mean curvature flow via modeling as a diffusion process, discretised in time with forward difference.
% \subsubsection{Negative volume formation}
% We use the Vatti clipping algorithm to obtain the residual shape of the bone, using the subtraction of intersected meshes. Boolean difference of two meshes results in the final volume of interest. 
% \begin{equation}
%     A \setminus B := \{ x \in {R}^3 \mid x \in A \textrm{ and } x \notin B \}
% \end{equation}
% \vspace{-0.2cm}

\subsection{Inflation \textit{vs.} 3D segmentation: why choose inflation?}
Supervised 3D segmentation models typically require extra labels to perform well. Given the time required to annotate our negative volumes manually ($\sim$1 hour, see Fig.~\ref{fig:pipeline_annot}), one would have to go through a very long annotation process to generate a proper dataset. Instead, we use lighter models for well-discernible bones and perform 3D inflation of the mesh, effectively mitigating the shortage of the labels and -- importantly -- also preserving the interpretability because the inflated volumes naturally `occupy' the available empty space in the joints.

Table~\ref{tab:table_loc} summarizes the key differences between the manual approach and the proposed automatic pipeline. Although our manual approach has a number of advantages over the clinical joint assessment methods, the machine-generated negative volumes are even better, being faster and entailing a more informative outer surface of the folume (see examples in Fig. 4 below and in the supplement).
% \begin{table}
% \caption{Summary of key characteristics between manual approach and our pipeline.}
% \label{tab:table_loc}
% \begin{center}
% \begin{tabular}{p{0.35\textwidth}p{0.15\textwidth}p{0.15\textwidth}p{0.15\textwidth}}
% \hline
% Feature & Clinical standard & Proposed \newline manual NV & Proposed  \newline automated NV & \\\hline
% Allows 3D measurements&   --& + & +\\
% Number of extracted parameters&   $\sim$2--16& $\sim$1--2$\times10^3$& $\sim$2--3$\times10^3$\\
% Defines exact anatomical shape& --& --& + \\
% Resilient to re-positioning& --& -- & +\\
% Hands-free report / Automation&   --& -- & + \\
% Segmentation Time&   0.5 hours & 1 hour & 4 seconds\\\hline
% \end{tabular}
% \end{center}
% \end{table}
\begin{table}
    \centering
    % {p{0.35\textwidth}p{0.15\textwidth}p{0.15\textwidth}p{0.15\textwidth}
    \caption{Summary of key characteristics between clinical and proposed methods.}
    \small
    \begin{tabular}{cccc}\toprule
    Feature & Clinical standard & Proposed manual NV & Proposed automated NV  \\ \midrule
    % Feature & Clinical standard & Proposed  & Proposed \\ & & manual NV & automated NV  \\ \midrule
    Allows 3D measurements&   --& + & +\\
    Number of extracted parameters&   $\sim$2--16 & $\sim$1--2$\times10^3$& $\sim$2--3$\times10^3$\\
    Defines exact anatomical shape& --& --& + \\
    Resilient to re-positioning& --& -- & +\\
    Hands-free report / Automation&   --& -- & + \\
    Segmentation time&   0.5 hours & 1 hour & 4 seconds\\ \bottomrule
    \end{tabular}
    {\small \raggedright \textbf{Note:} NV stands for the negative volume. \par}
    % \caption{Caption}
    % \label{tab:my_label}
    \label{tab:table_loc}
\end{table}
\section{Experiments}
\label{sec:experiments}
% \vspace{-0.2cm}
\subsection{Dataset}
% \vspace{-0.2cm}
\label{sec:dataset}
% \subsubsection{DICOM and STL preprocessing}
Our dataset contained 50 patient’s head CT scans
% , acquired on a Planmeca ProMax scanner with spatial 
with the resolution of 0.4 mm and the dimensions of
% . Every scan consists of voxels with size of approximately 
$686 \times 686 \times 686$ pixels. 
% Although MRI is thought to be the gold standard for TMJ visualization, CT is capable of showing detailed bone structures that is particularly necessary for bone tissue segmentation. 
% Thus, CT is more appropriate imaging modality for our task. 
The ground truth masks [20 STL models of 10 patient's mandibular heads (i.e., left and right TMJs)] were obtained after the manual annotation by two experienced orthodontists following the pipeline shown in Fig.~\ref{fig:pipeline_annot} in the MIMICS program.
% using threshold segmentation combined with manual corrections. 
% labels of all 20 STL models (left and right TMJ of each patient) 
The STL models were voxelized by the subdividing method: a mesh was scaled down until every edge was shorter than the spatial resolution.

The train test split was done by patient id, as it is a standard for medical datasets. To avoid overfitting, all models were trained using 5-fold cross-validation on 10 patients with annotated masks. This was made to have all available labeled data in the training group, thus, increasing the accuracy for the remaining 40 patients in the hold-out test. To further minimize the overfitting problem originating from the limited training set, we applied a large variety of data augmentation techniques: random 3D rotation, horizontal flipping, contrast, translation, and elastic deformations. All the augmentation techniques were applied on the fly during training.

\subsection{Training of the Neural Network}
% \vspace{-0.2cm}
The pipeline is implemented using Pytorch\footnote{Experiments were conducted on a server running Ubuntu 16.04 (32 GB RAM); the training was done on NVidia Geforce Ti 1080 GPU (11 GB RAM).}. In all experiments, we use a 5-fold cross-validation and report the mean performance. 

\textbf{TMJ localization.}
For localization training, $160\times 160\times 160$ images and a combination of both masks (TB and MC) are used with a batch size of 1 for memory considerations. We use Adam optimizer with learning rate 0.001 and parameters $\beta_1 = 0.9$, $\beta_2 = 0.99$. 
The weight decay regularization parameter is equal 0.01. 
% The learning rate decrease with a factor of 10 on epoch 20 and 40. We use an early-stop mechanism and train till the validation loss stops decreasing for 10 epochs in a row.
% It took an average of 35 epochs (1400 iterations) to meet this criterion. 
Linear combination of Cross-Entropy (CE) and Dice loss was used as a loss function to optimize both a pixel-wise and overall quality of segmentation. 
% While Mean Square Error (MSE) between corner points of bounding boxes for labels and output data was used as evaluation metrics of localization results. 
After obtaining a rough segmentation of the joint area, automatic postprocessing was performed, including thresholding based on the minimum method and morphological operations to remove outliers. 
% We show how this step results in defining the borders for VOI in the supplementary material.
%Fig.~\ref{fig:localization}

\textbf{MC and TB segmentation.}
The segmentation models are trained on $112\times 144\times 64$  patches form resulted VOIs, which differ slightly on all scans. Adam optimizer is used with initial learning rate of 0.0001. 
% Similar to localization task the learning rate is divided by 10 at the 30th, 60th, and 90th epochs. 
Each model is trained for 100 epochs (8000 iterations) to ensure convergence. We did not perform specific hyperparameter tuning and used fixed hyperparameters for an honest comparison.
%In addition to original V-Net architecture, we added drop-out in middle levels to improve generalization ability. (???)
We run the training with Cross-Entropy (CE), Dice loss (D), or their linear combination to evaluate the impact of these metrics on segmentation performance. Dice score (DICE), Cross-Entropy, and Hausdorff distance (HD) were used to evaluate the performance of segmentation.

\section{Results}
\paragraph{Joint localization}
The V-Net model used for localization task reached the Dice coefficient $64.6 \pm 0.3 \%$ and
Cross-Entropy $0.040 \pm 0.001$ for evaluation of coarse segmentation on full CT scans and MSE is $7.940 \pm 2.009$ for determination of bounding boxes around joints. We show the visual results of localization together with the resulting VOI boundary in the supplementary material.
%Fig.~\ref{fig:localization} 
It confirms that the achieved quality is sufficient to approximate the location of the joint, since the in the collected dataset, as well as in general clinical practice, there is no single way to determine the exact boundaries of the joint.
% MULTI-HEAD fig is now in supplement
% \vspace{-0.25cm}
\paragraph{Mandibular condyle and temporal bone segmentation}
\label{sec:results}
% \vspace{-0.25cm}
Table~\ref{tab:table_segmentation} shows the results of the 3D U-Net, 3D U-Net with attention, and V-Net models trained with different loss functions for the bone segmentation blocks in Fig.~\ref{fig:pipelie_Vnet}. 
% It justifies selection of V-Net architecture, which perform best for segmenting MC in terms of all chosen metrics and for segmenting TB with D + CE. 
It justifies selection of V-Net architecture trained with D+CE, which perform best for segmenting MC in terms of all chosen metrics and achieves an average Dice score of 91.4 \% and Cross-Entropy of 0.154, which is of the state-of-the-art level in various well-annotated  segmentation reports~\cite{VNET_MICCAI18_WORKSHOP,DLReview_MED}. For TB segmentation, V-Net also outperform 3D U-Net and 3D~U-Net with attention in terms of HD and it is not much inferior in other metrics. 
% Unlike those studies, though, 
We note that the TB annotation can very rough at best due to such a complex shape of this bone, making it very hard to gauge segmentation performance by simple comparison to the ground truth labels (see the supplemental material for visual assessment and Fig.~\ref{fig:shapes} below).
The relatively high values of the Hausdorff distance in Table~\ref{tab:table_segmentation} support this notion and reinforce the idea behind the auxiliary classical processing (step C in Fig.~\ref{fig:pipelie_Vnet}) required for the insufficiently annotated datasets.
\begin{table}
    \centering
    \caption{Mandibular condyle (MC), temporal bone (TB) and negative volume (NV) segmentation results. Notice that the whole-object 3D segmentation of the manually annotated ``balls'' from Fig.1 need more data to work properly, justifying the development of our automated pipeline which just needs MC and TB masks.}
    \small
    \begin{tabular}{ccccccc} \toprule
    % \multicolumn{2}{c}{Item} \\ \cmidrule(r){1-2}
    Obj. & Score & 3D U-Net & 3D U-Net+Att.  &  V-Net CE & V-Net D & V-Net D+CE \\ \midrule
    \multirow{3}{*}{MC}
    %   \multirow{3}{*}{MC}
     & DICE & $91.4 \pm 5.3$ & $89.8 \pm 8.2$ 
     & $90.9 \pm 4.5$ & $90.9 \pm 6.3$  & $\mathbf{91.4} \pm 4.8$ \\ % VNet 
     & CE & $0.320 \pm 0.003$ & $0.320 \pm 0.005$ 
     & $0.201 \pm 0.075$ & $0.175 \pm 0.024$ & $\mathbf{0.154} \pm 0.053$ \\ % VNet  
     & HD & $14.7 \pm 20.8$ & $15.2 \pm 21.6$ 
     & $11.9 \pm 15.7$ & $11.5 \pm 20.1$  & $\mathbf{10.5} \pm 21.2$  \\ \cline{1-7} % VNet 
       
    \multirow{3}{*}{TB}
     & DICE & $75.5 \pm 8.8$ & $75.8 \pm 8.4$ 
     & $75.9 \pm 6.9$ & $\mathbf{76.7} \pm 6.8$  & $76.3 \pm 7.2$ \\ % VNet 
     & CE & $0.463 \pm 0.043$ & $0.462 \pm 0.035$ 
     & $\mathbf{0.383} \pm 0.088$ & $0.396 \pm 0.093$ & $0.416 \pm 0.100$ \\ % VNet  
     & HD & $29.8 \pm 11.5$ & $29.9 \pm 11.3$
     & $27.9 \pm 11.5$ & $ 28.3 \pm 10.7$ & $ \mathbf{27.6} \pm 10.9$ \\ \cline{1-7} % VNet 
     
    \multirow{3}{*}{NV}
     & DICE & $78.0 \pm 10.6$ & - & - & - & $77.7 \pm 7.7$\\ %\cline{2-7} 
     & CE & $0.344 \pm 0.016$ & - & - & - & $0.406 \pm 0.022$ \\ %\cline{2-7} 
     & HD & $15.8 \pm 18.8$ & - & - & - & $18.7 \pm 17.8$ \\ \bottomrule
    \end{tabular}
     {\small \raggedright \textbf{Note:} Here \textit{CE}, \textit{D} are Cross-Entropy and Dice loss, respectively. \textit{DICE} (measured in $\%$) and \textit{HD} are Dice score and Hausdorff distance. \textit{Att.} stands for the attention-gate architecture.  \par}
    \label{tab:table_segmentation}
\end{table}
% \begin{table}[]
% \caption{Another variant of table above.\hl{YES! Let's go ahead with this one}}
% \label{table_segmentation_NIPS}
% \scriptsize
% % \begin{center}
% \begin{tabular}{|l|l||l|l|l|l|l|}
% \hline
% % \multirow{}{}{Approach} & \multicolumn{3}{c||}{Mandibular condyle (MC)} & \multicolumn{3}{c|}{Temporal bone (TB)} 
% % \\ \cline{2-7} 
% Object & Metric & UNet 3D & UNet 3D + Attn.  &  VNet, CE loss & VNet, Dice loss & VNet, Dice + CE  \\ \hline\hline
% \multirow{}{}{MC}
%  & Dice, \% & 0 & 0 & 0 & 0 & 0\\ %\cline{2-7} 
%  & Cross Entropy & 0 & 0 & 0 & 0 & 0 \\ %\cline{2-7} 
%  & Hausdorff Distance & 0 & 0 & 0 & 0 & 0\\ \cline{1-7}
   
% \multirow{}{}{TB}
%  & Dice, \% & 0 & 0 & 0 & 0 & 0\\ %\cline{2-7} 
%  & Cross Entropy & 0 & 0 & 0 & 0 & 0 \\ %\cline{2-7} 
%  & Hausdorff Distance & 0 & 0 & 0 & 0 & 0\\ \cline{1-7}
% \multirow{}{}{Neg. Vol.}
%  & Dice, \% & 0 & - & - & - & 0\\ %\cline{2-7} 
%  & Cross Entropy & 0 & - & - & - & 0 \\ %\cline{2-7} 
%  & Hausdorff Distance & 0 & - & - & - & 0\\ \cline{1-7}
% \end{tabular}
% % \end{center}
% \end{table}
% According to comparison of segmentation results for TB, we observe that V-Net outperforms U-Net based architectures only in terms of cross-entropy, while 3D U-Net demonstrates the best results in Dice metrics and Hausdorff distance.
% \vspace{-1cm}
\begin{figure}[b]
    \centering
    % Thin
    % \includegraphics[width=1\textwidth]{EPS/VEC_fig4_upd_compared_thinner-01.pdf}
    % Normal below ->>>>>>>>>>>>>>
    % \includegraphics[width=1\textwidth]{EPS/VEC_fig4_upd_compared_ok.pdf}
    \includegraphics[width=1\textwidth]{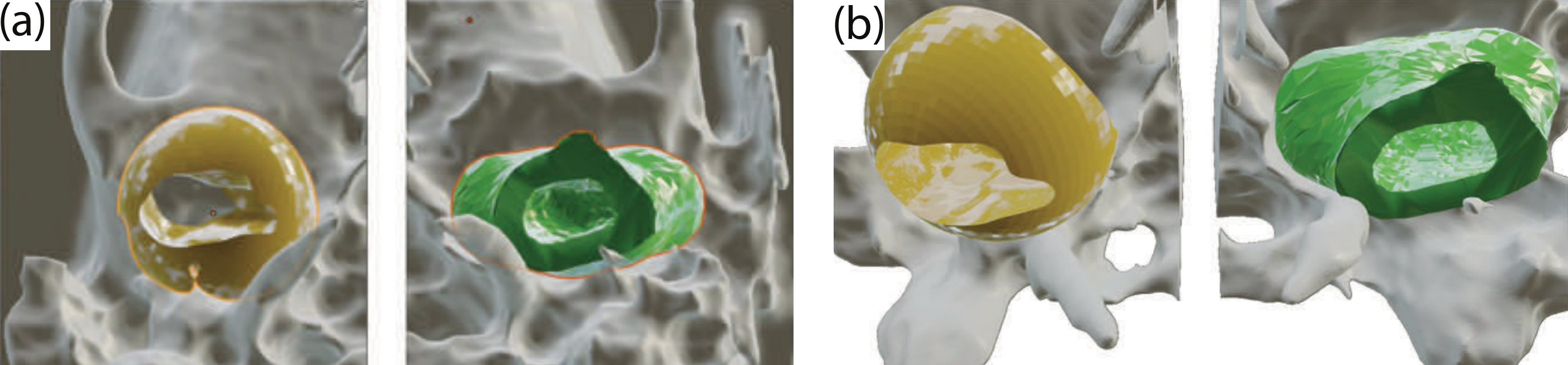}
    \caption{Proposed manually annotated (yellow) \textit{vs.} machine-generated (green) negative volumes. Rendered regions of the TB are shown in gray. Views: (a) axial, from bottom (b) same, tilted.}
    \label{fig:shapes}
\end{figure}
\paragraph{Machine-found negative volumes}
3D-reconstructed volumes of the segmented MC bones are then ``inflated'' as shown in Fig.~\ref{fig:inflation}. The result of such operation for a single patient is shown in Fig.~\ref{fig:shapes}, which compares the manually annotated negative ``ball'' (yellow, pipeline of Fig.~\ref{fig:pipeline_annot}) and the non-spherical machine-generated negative volume (green, pipeline of Fig.~\ref{fig:pipelie_Vnet}).
Remarkably, despite being much more informative than the linear measurements, our manual annotation solution still struggles to portray the full complexity of the ``negative space'' in the joint. On the contrary, the machine-generated negative volumes effortlessly occupy the space available within the joint and, thus, summize \emph{complete volumetric characterization} of the joint. Our end-to-end algorithm generates such volumes $\sim$100-fold faster than the human, taking about 4 seconds to compute.
 
\begin{figure}[b]
    \centering
    \includegraphics[width=1.0\textwidth]{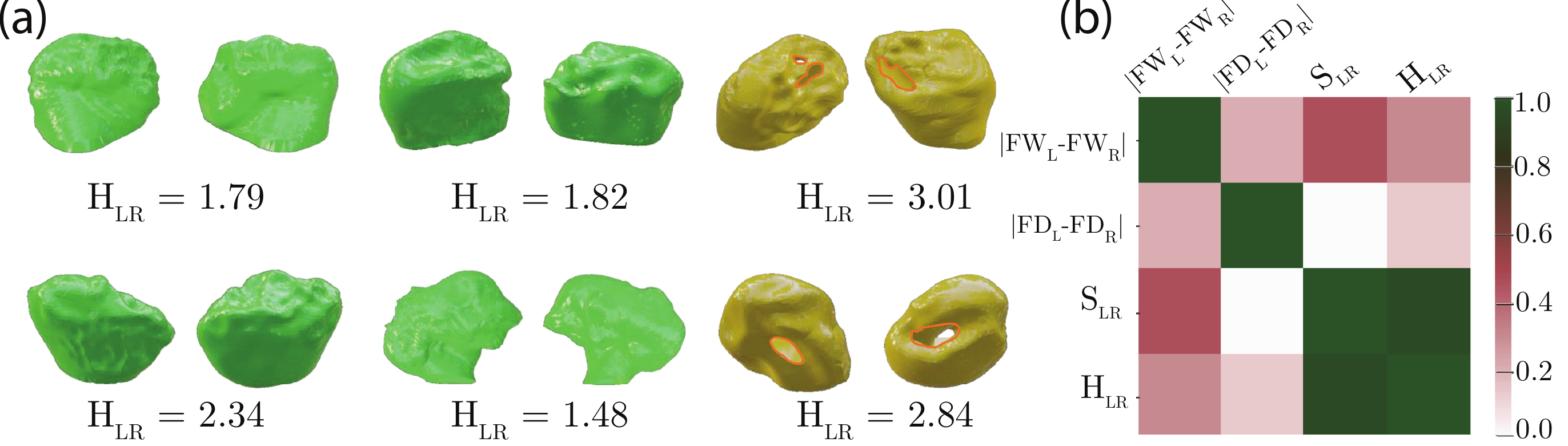}
    \caption{(a) Negative volumes of 6 patients from the Table~\ref{tab:6patients} and their symmetry metrics. Notice unevenly worn out joints in the last column (TMD patients). (b) Correlation between the proposed and the state-of-the-art symmetry measures for the entire dataset.}
    \label{fig:6patients}
\end{figure}

We generated pairs of negative volumes for all 50 patients, and showed measurements for six of them in Fig.~\ref{fig:6patients} and in Table~\ref{tab:6patients}. 
Although rudimentary, the clinical measurements correlate with the proposed volumetric metrics in the task of detecting the worn joints (see 50-patient heatmap in Fig.~\ref{fig:6patients}(b)), implying that the new volumetric metrics $S_{LR}$ and $H_{LR}$ could be proposed for adoption to the current practices of the maxillofacial medicine. 

% NeurIPS recommended styles - booktabs package
\begin{table}
     \caption{Proposed negative volume symmetry metrics $S_{LR}$ and $H_{LR}$, and the rudimentary linear measurements currently used in clinics (''FW'' and ''FD'' marked in Fig.~\ref{fig:pipeline_annot}).}\label{nv_measures}
    \centering
    \begin{tabular}{ccccccc} \toprule
    % \multicolumn{2}{c}{Item} \\ \cmidrule(r){1-2}
    Patient & $FW_{L}$, mm & $FD_{L}$, mm & $FW_{R}$, mm & $FD_{R}$, mm &  $S_{LR}$ & $H_{LR}$ \\ \midrule
    1 & 15.6  & 6.8  & 15.2  & 6.7  & 1.02 & 1.79 $\pm$ 0.25 \\  
      2 & 14.6  & 6.3  & 16.4  & 7.2  & 1.03  & 1.82 $\pm$ 0.28 \\ 
      3 & 17.5  & 7.3  & 16.7  & 7.4  & 1.02 & 1.48 $\pm$ 0.31\\ 
      4 & 18.3  & 7.9  & 18.9  & 7.5  & 1.15 & 2.34 $\pm$ 0.29 \\
      5 & 16.6  & 7.7  & 21.2  & 6.8  & \bfseries{1.17}  & \bfseries{2.84} $\pm$ 0.27 \\ 
      6 & 16.3  & 6.8  & 19.8  & 6.7  & \bfseries{1.21} & \bfseries{3.01} $\pm$ 0.28  \\ \bottomrule
    \end{tabular}
    \label{tab:6patients}
\end{table}
\section{Discussion and Conclusions}
Notice that the patients with confirmed jaw misalignment (patients no. 5 and 6 in Table~\ref{tab:6patients} and Fig.~\ref{fig:6patients}(a)) have distinct pathological profiles in the negative volumes. These cases emphasize how important it is to have the full volumetric representation of the empty space within the joint. What could also be concluded from Table~\ref{tab:6patients}, is that the proposed metrics are not exclusive: we observe that $S_{LR}$ is more specific and is better suited for large asymmetry, whereas $H_{LR}$ is more sensitive to miniature differences in the shape, such as those in the TB bone. Modern topological metrics, e.g. Wassertein distance, could further enhance asymmetry detection by taking advantage of the optimal transport theory~\cite{wasserstein_dist}. Another line of future work calls for continuation of data collection and annotation. We publish our end-to-end pipeline\footnote{ https://github.com/cviaai/DEEP-NEGATIVE-VOLUME.} and can envision its seamless integration into active learning tools to alleviate the annotation burden.

To summarize, we proposed a new intuitive hybrid strategy for medical 3D image segmentation, entailing new manual annotation pipeline, localization-based image enhancement, deep learning-based segmentation, and surface mesh inflation. 
The framework extracts ``negative volumes'' in complex anatomical structures in an end-to-end manner, which we validated on a head-CT dataset by segmenting the most complex human joint (the TMJ) together with maxillofacial experts. Our method is two orders of magnitude faster than the manual segmentation and as much more informative compared to the current practices. We, therefore, propose this method as a new joint health assessment technique for the large cohort validation and consequent clinical adoption.
% \begin{ack}
% Use unnumbered first level headings for the acknowledgments. All acknowledgments
% More information about this disclosure can be found at: \url{https://neurips.cc/Conferences/2020/PaperInformation/FundingDisclosure}.

% Do {\bf not} include this section in the anonymized submission, only in the final paper. You can use the \texttt{ack} environment provided in the style file to autmoatically hide this section in the anonymized submission.
% \end{ack}

% \include{impact}
\section*{Broader Impact}

Modern computer vision software shows impressive accomplishments in extracting and understanding a plethora of 3D object shapes from various imaging applications. Following the advent of deep learning (DL), the segmentation of 3D objects could be now done with excellent quality. Yet, the segmentation of the truly intricate compound 3D objects still remains an essential challenge. We propose an elegant and an intuitive approach to avoid the hard-to-annotate regions of a compound 3D object, and -- instead -- learn how to segment `the air' within the 3D object of interest. We coined this `air' as a "Negative Volume" and proposed the first DL framework for segmenting them automatically.

In this work, we showed segmentation of a particularly complex joint in the human jaw (allegedly, the most complex one in the body). The method, however, is universal, and the methodology of DL-based segmentation of negative volumes could impact disciplines beyond healthcare, ranging from the additive manufacturing, to the seismic sensor 3D data, to detecting underground objects in oil and gas, to extracting complex scenes from LIDAR data in self-driving cars.

% \paragraph{Healthcare applications -- human joints and beyond} 
% The main paper concerns one of the most complex joints -- TMJ. Amongst the most important body parts, knee caps are also most affected by mechanical deformations during a person's lifespan. Increasing use of high-tech prostheses requires increasing precision of preparatory actions. As such, accurate and robust identification of the damaged body joints may not just decrease the costs of such production, but also have a significant impact on the patient's life quality in the long-term perspective.

% \paragraph{From Healthcare to Other Industries}
% One of the crucial step in the interpretation of seismic 3D images, main source of data for analysis of Earth’s subsurface, is the separation of such seismic cube into different layers of rocks (strata). Our approach aimed to definition of volumes, where clear boundaries are not present, can become a new way to delineate the boundaries between different seismic structures, effectively estimating the volume of contained resources, such as oil, natural gas, salt, and water from the earth's interior.
% \section{Broader Impact}
% Two paragraphs!
\newpage

\section{Supplementary Material}

Our Supplementary material is structured as follows. First, in Section~\ref{s:dataset}, we present details about the dataset, demonstrating the anatomical diversity and shape complexity of different TMJ joints. In Section~\ref{s:nv}, we illustrate the major difference between the manually annotated negative volumes and the ones directly segmented given that 3D annotation. In Section~\ref{s:segm_res}, we then discuss the segmentation performance as a function of the training set size and of the training loss function, given the metrics considered in the main text. Last, we provide in-depth medical literature review, omitted from the main text in Section~\ref{s:joint_literature} to describe the \emph{clinical} state-of-the-art in the area of the human joint space assessment.
% In Section~\ref{s:local} we present the localization results for a CT scan in different planes. We show the effect of the training size by comparison of our models' performance graphs featuring dependency between the localization results on the test set and the size of the training set. Then

\begin{figure}[h]
    \centering
    \includegraphics[width=1\textwidth]{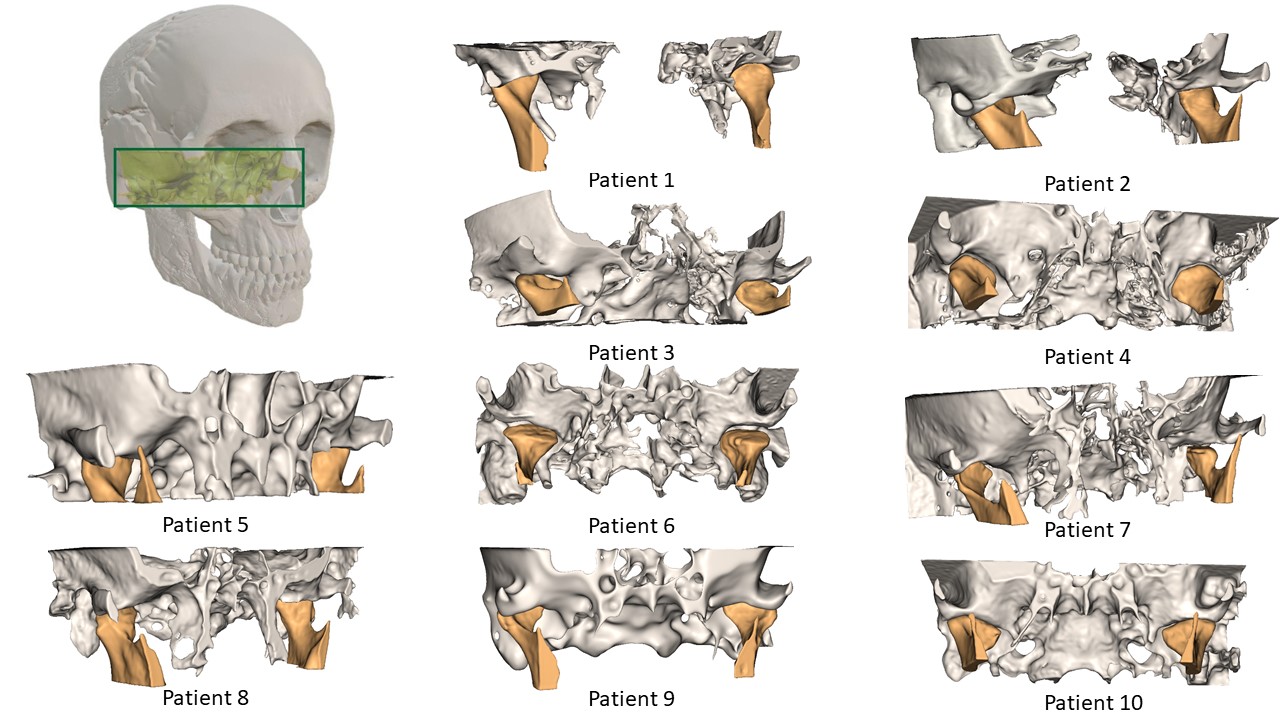}
    \caption{Anatomical diversity of TMJ consisting pf the mandibular condyle (orange) and temporal bone (gray) for 10 patients from the dataset. For 1 and 2 patients two volumes of interest (VOIs), containing left and right joint, were individually selected. While for patients 3-10 only one VOI, bounding two joints, was chosen, which results in one STL model for both temporal bones (left and right).}
    \label{fig:tmjs}
\end{figure}

\section{Dataset}\label{s:dataset}
To validate our deep negative volume segmentation approach, we use a local dataset containing DICOM scans of the heads of 50 patients. The dataset was acquired in a Pavlov First St.Petersburg State Medical University hospital\footnote{The dataset will be made public after the review.}. Two experts in maxillofacial medicine annotated the temporal mandibular joint (TMJ) using the proposed annotation workflow (see Fig. 1 in the main text). 

Fig.~\ref{fig:tmjs} shows the location of TMJ in a skull and demonstrates variability in the joint area definition, as well as anatomical diversity of TMJ shapes. It is assumed that symmetry should exist between contralateral sides in the same individual.

\begin{figure}[h]
    \centering
    \includegraphics[width=1\textwidth]{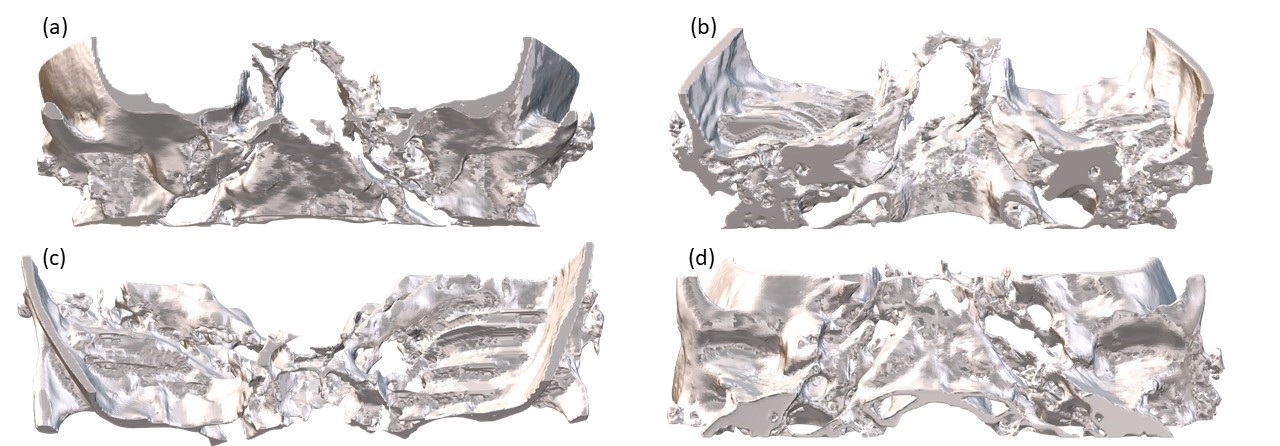}
    \caption{One model for both temporal bone (left and right) in different views: coronal from the front (a), coronal from the back (b), axial from the top (c), axial from the bottom (d).}
    \label{fig:TB_complexity}
\end{figure}

Despite the fact, that all TMJ components vary considerably both in size and shape, mandibular condyle has a simple recognizable form, resembling an oval from above, while temporal bone has a much more complicated configuration due to the plenty of spikes and irregularities. Fig.~\ref{fig:TB_complexity} illustrates how complex the temporal bone structure is. 

\section{Negative Volume Segmentation}\label{s:nv}
\begin{figure}[h]
    \centering
    \includegraphics[width=1\textwidth]{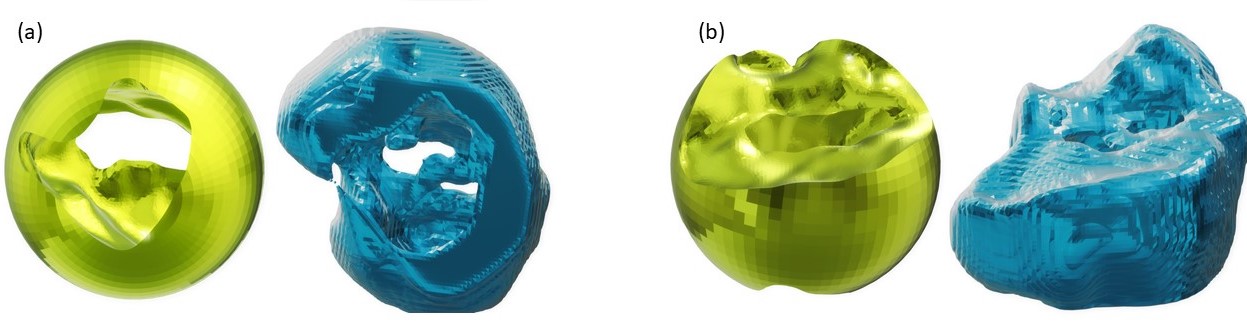}
    \caption{Comparison of a manually annotated negative volume (green) and a reconstructed one after segmentation (blue) from the bottom (a) and from the side viewangle (b). The figure demonstrates that a straightforward segmentation of the negative volume is not capable of extracting empty space withing a spherical object well, especially if only a limited annotation is available.}
    \label{fig:segm_ball}
\end{figure}

In order to extract the necessary negative volume, we first tried to implement its segmentation directly in a supervised learning manner. It proved to be a difficult task, since the outer boundaries of the negative volume represent an almost perfect sphere and are not anatomically defined. Fig.~\ref{fig:segm_ball} shows the result of this segmentation and compare manually annotated negative volume with reconstructed one. Fig.~\ref{fig:B_segm} demonstrates in 2D axial slices how poorly the models try to repeat the round non-anatomical contour of the annotated "ball".

\section{Localization Results}\label{s:local}
% related work
Architectures based on the Convolutional Neural Networks and related to localization of compound joints i.e. such as knee~\cite{knee_segmentation, knee_box} and hips~\cite{segmentation_by_detection} are of special interest for our task, since the aim is to localize junction between bones (small adjoining parts of the bones) and not the object entirely as in case of organs or tumors. These approaches aimed to automatically detect joint region utilize a coarse segmentation to solve the localization problem. 

% importance
Similarly, the localization step is essential in our case of joint structures segmentation, since we have high-resolution input ($686\times 686 \times 686$) with small area of interest and inconsistent joint annotation in the sense that the original joint VOI was chosen not strictly anatomically but intuitively.

\begin{figure}
    \centering
    \includegraphics[width=1\textwidth]{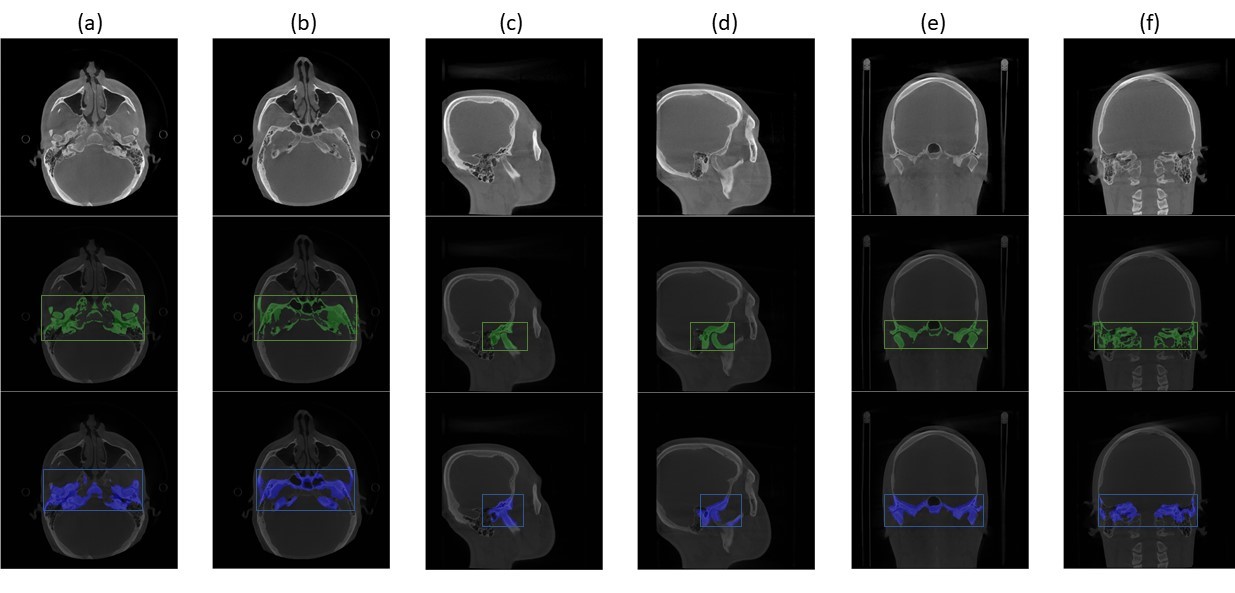}
    \caption{Automatic VOI localization results: (a), (b) axial view; (c), (d) sagittal view; and (e), (f) coronal view. Top row: raw images. Middle row: ground truth bounding box (green). Bottom row: automatic localization output (blue).}
    \label{fig:localization}
\end{figure}

% results
To perform localization of joint we utilize V-Net architecture on full CT scans and treat it as a problem of coarse segmentation at a lower resolution. Fig.~\ref{fig:localization} represent visual results of localization together with the bounding VOI for both joints in different planes. In order to facilitate further segmentation, we crop resulted VOIs into 3 parts by sagittal cross sections in a such way that right and left parts contain joints, and the middle one does not. 
Thus, for each CT scan localization step results in two cropped volumes which are on average equal to $144\times  150\times  117$ but differ slightly for all patients.

\section{Segmentation Performance}\label{s:segm_res}

\subsection{Influence of the training set size}\label{ss:train_size}

We investigate how the size of training set affects bone segmentation performance. Since only 10 annotated scans were available, we evaluate the performance of models in cross-validation procedure, increasing the number of patients in training set from 1 to 9. For the size of training set from 1 to 8 we perform 5-fold cross-validation, as for the experiments in the main part, and for the training size of 10 we use leave-one-out cross-validation.

\begin{figure}
    \centering
    \includegraphics[width=1\textwidth]{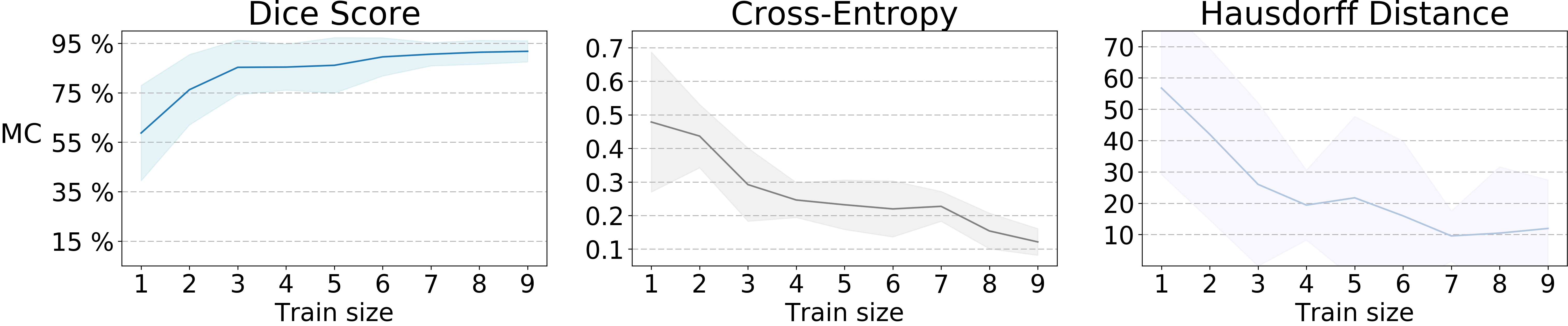}
    \includegraphics[width=1\textwidth]{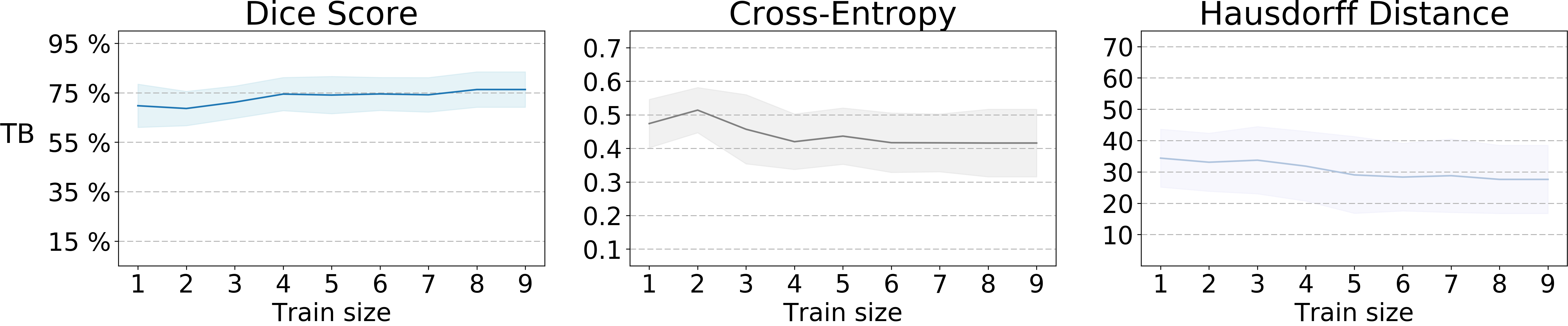}
    \caption{Dependence of performance on the test set and on the size of the training set (the number of patients) for MC and TB segmentation in the first and the second row, respectively.}
    \label{fig:training_size}
\end{figure}

Fig.~\ref{fig:training_size} demonstrate how the size of the training set affects segmentation performance. Although the performance gain from the additional training data tends to decrease, extending the dataset can stably improves results. This is especially noticeable for MC, where Dice score of segmentation reaches 92\% with 9-patient training set, while for TB performance growth looks less significant and  more linear.
According to our observations, when segmenting such complex structures the quality of annotation is no less important than the quantity. Future work could study how these approaches perform when more data is available.

\subsection{Influence of the training loss}\label{ss:train_loss}
\begin{figure}
    \centering
    \includegraphics[width=1\textwidth]{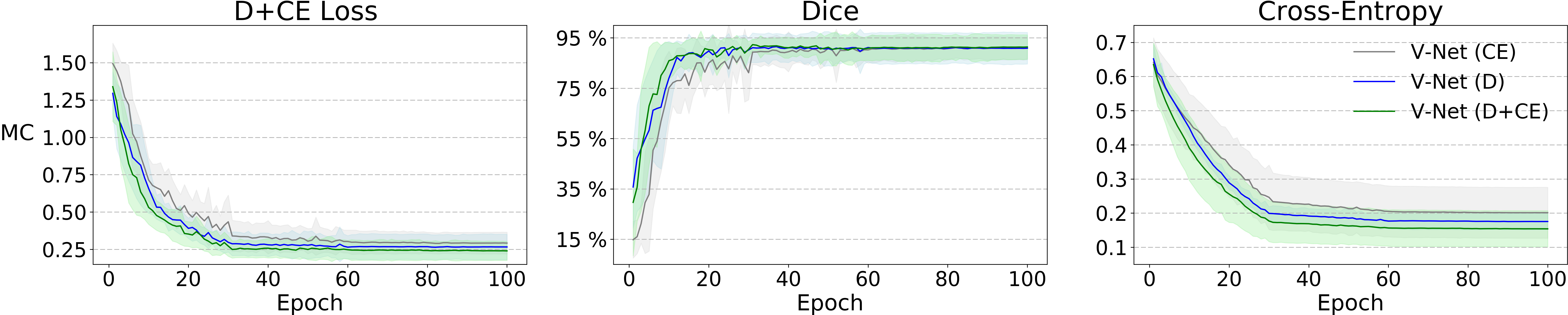}
    \includegraphics[width=1\textwidth]{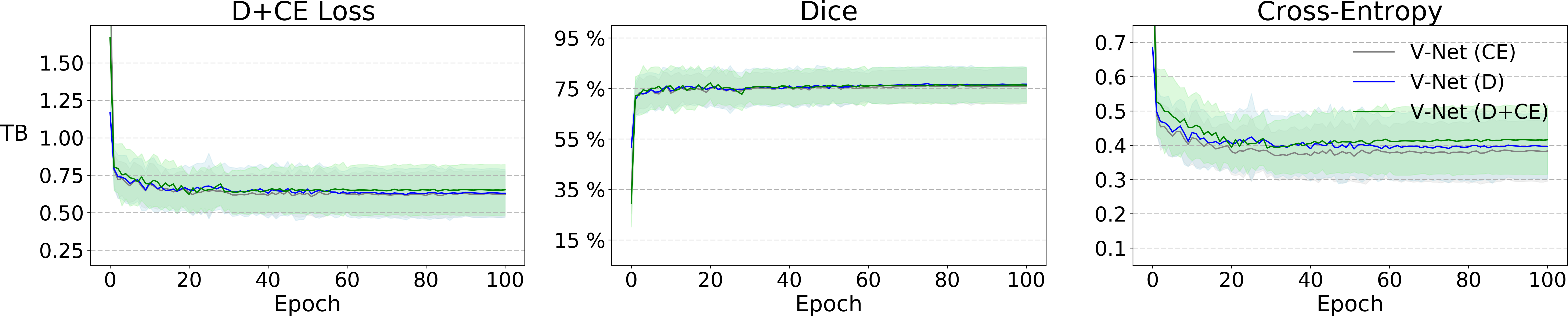}
    \caption{Comparison of V-Net convergence on validation tests with different training loss: Cross-Entropy (CE), Dice loss (D), and their combination (D+CE).}
    \label{fig:vnet_conv}
\end{figure}

We compare the performances of V-Net architecture training with different loss functions: Cross-Entropy (CE), Dice loss (D), and their linear combination (D+CE) to evaluate the impact of these metrics on segmentation results. Fig.~\ref{fig:vnet_conv} demonstrate convergence of these loss functions. 
For MC segmentation, the convergence rate with D+CE loss is noticeably faster for both Dice score and Cross-Entropy and resulted Cross-Entropy value is smaller.
While for TB, the Dice score is almost independent of the loss selection, and with CE loss Cross-Entropy value is, as expected, slightly better. Taking into account the superiority of D+CE loss in Hausdorff distance (see Table 2 in the main part) and overall visual comparison of reconstructed bones in Fig.~\ref{fig:segm_3d}, we chose model configuration with D+CE loss for further steps. 

\subsection{Visual results}
We provide qualitative examples to assess visually the behavior of segmentation models in 2D axial slices for MC in Fig.~\ref{fig:M_segm} and TB in Fig.~\ref{fig:TB_segm}, as well as in 3D reconstructed outputs for both bones in Fig.~\ref{fig:segm_3d}  

\begin{figure}
    \centering
    \includegraphics[width=1\textwidth]{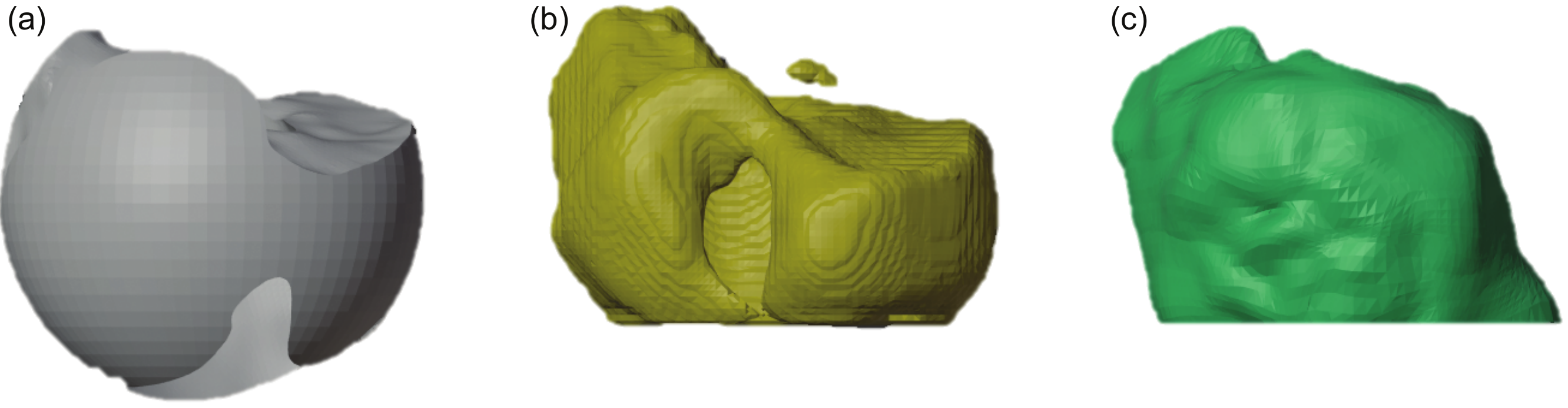}
    \caption{Side view of the negative volumes (Patient 8) generated by: (a) manual annotation; (b) 3D segmentation given the manual annotation; (c) bone segmentation followed by the inflation procedure, as discussed in the main text. Notice that the manual annotation entails the idealistic sphere which misses important details of the true morphology within the joint. Such an idealistic annotation could be used to train a segmentation model, as shown in (b), however, the model fails to learn how to 'fill' the empty space between the bones. To the opposite, our proposed inflation method eliminates the problem and requires much less data to train the neural network (Please see the supplementary video for a better view: \texttt{negative\_volumes.mp4}).}
    \label{fig:compared}
\end{figure}

\begin{figure}
    \centering
    \includegraphics[width=1\textwidth]{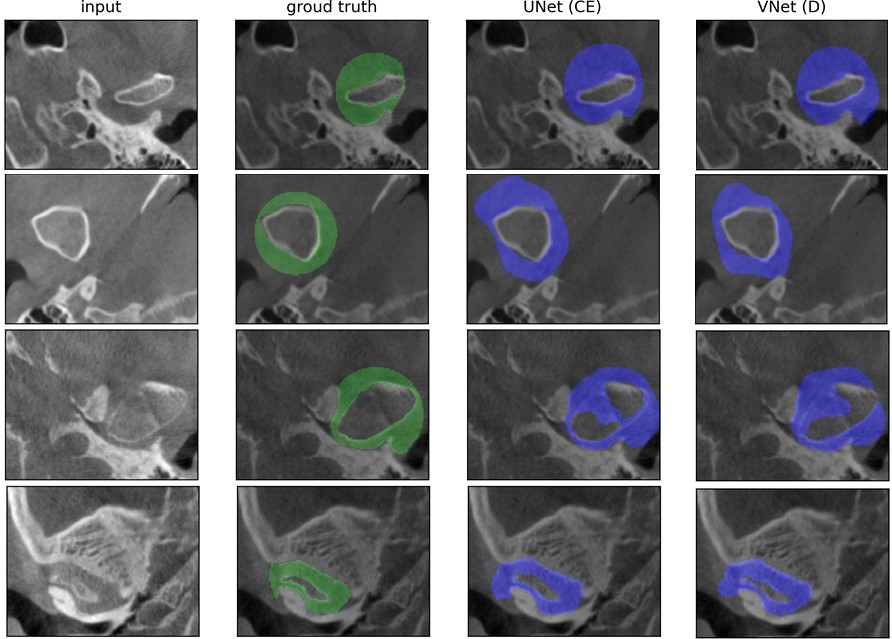}
    \caption{Segmentation results for spherical negative volumes. Note how fruitlessly the models try to repeat the round non-anatomical contour of the "ball".}
    \label{fig:B_segm}
\end{figure}

\begin{figure}
    \centering
    \includegraphics[width=1\textwidth]{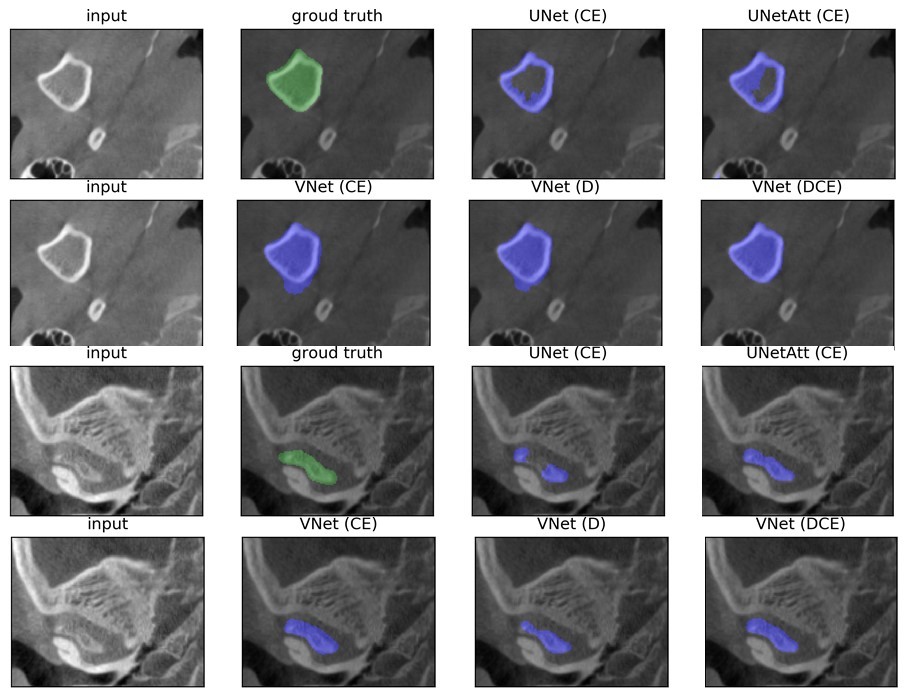}
    \caption{Segmentation results for the most difficult examples of mandibular condyle (MC): the first input (first two rows) is different in shape from the typical  examples of MC represented in the training set, and the second input (last two rows) has unclear boundary of bone tissues. Our inflation pipeline allows to mitigate and to generate the negative volumes even for these patients.}
    \label{fig:M_segm}
\end{figure}

\begin{figure}
    \centering
    \includegraphics[width=1\textwidth]{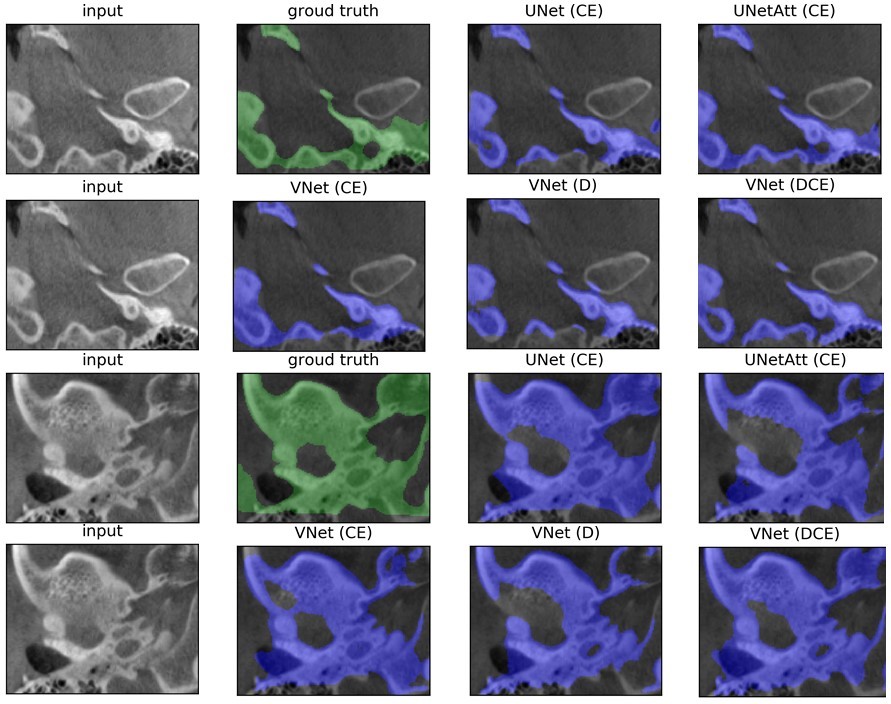}
    \caption{Segmentation results for temporal bone. Note the fuzzy boundary of the temporal bone even in ground truth (green), that presents hard to annotate area.}
    \label{fig:TB_segm}
\end{figure}

\begin{figure}
    \centering
    \includegraphics[width=1\textwidth]{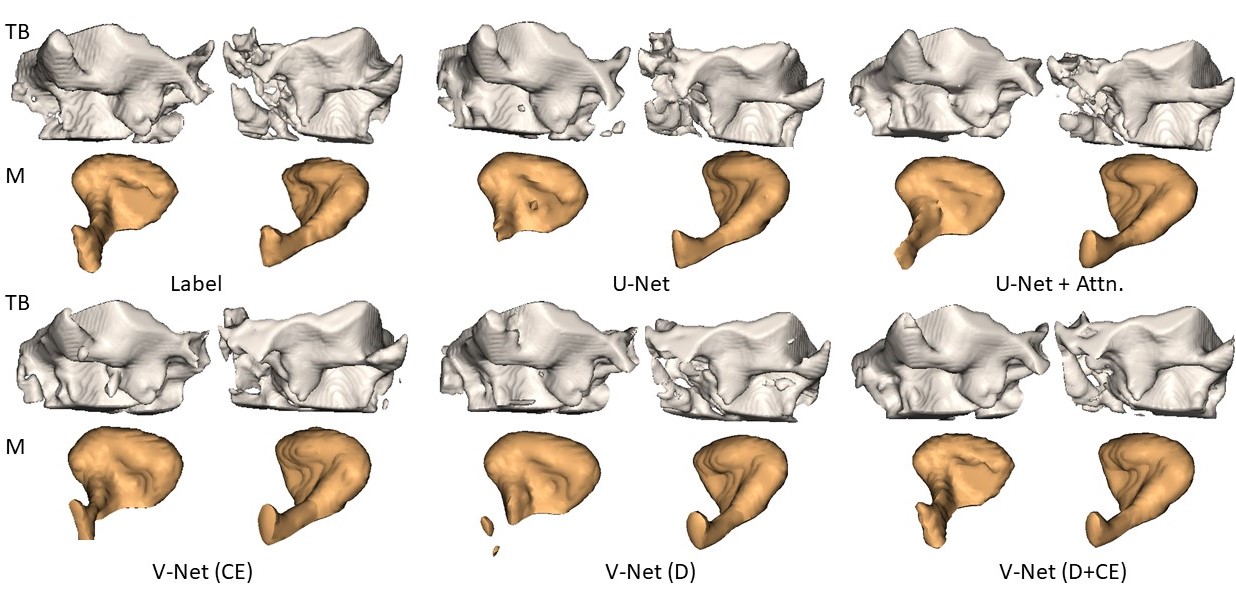}
    \caption{Results of TMJs segmentation for one patient using 3D U-Net, 3D U-Net with attention, and V-Net architectures. Ground truth labels and reconstructed models for bone components of TMJ: temporal bone (TB, gray) and mandibular condyle (MC, orange) are shown.}
    \label{fig:segm_3d}
\end{figure}

\section{Literature review on joint space assessment}\label{s:joint_literature}
% intro
This section is a supplementary of Section 2 in the paper. It highlights the clinical significance of joint space assessment and describes existing methods in this area. Special attention is paid to TMJ as the most complex joint and the main object of this research. 

% Metacarpophalangeal joint space
One group of joints, for which the assessment of the joint space is of particular value, are the metacarpophalangeal joints located between the metacarpal bones and the proximal phalanges of the fingers. 
According to the review~\cite{wrist_review} of the current methods of joint space evaluation for wrist joints, three approaches had significant developments for fully automated, quantitative 3D measurements of width and volume of 3D joint space~\cite{jsw1,jsw2,jsw3}. 
While some of these techniques have demonstrated high-throughput, robust, reproducible capacity required in medical practice, there is a discordance in the definition of the joint space volume of interest between the three algorithms. All of them are based on applying erosion/dilation morphological operations for joint space detection and then apply a distance transform algorithm, which in essence fits spheres in the masked volume of interest to obtain a measure of width. However, the side borders of the resulted joint space are highly dependent on the initial parameters of erosion/dilation morphological operations, which leads to different calculations of three-dimensional metrics. One research collaboration made an extensive comparison of these methods to know whether or not data can be interchangeably analyzed by any of these methods. In addition, a consensus approach for evaluation of 3D joint space width was proposed, ensuring large spheres near the border of the joint space volume mask are not excluded and reducing dependency on the parameters of dilations used in the closing operation~\cite{metacarpophalangeal}.
Despite the fact the suggested hybrid method reduces the differences between evaluations and ensure reproducibility and ease of use, it does not solve the problem of determining the clear boundaries of the joint space. In essence, conducted research highlights the need for a solution to handle the "border issues" of joint space volume.

% TMJ space
The literature review devoting to TMJ space~\cite{tmj_space} 
suggests that although many studies were conducted on articulate space, the results were highly individualized and incomparable due to the high heterogeneity in terms of sample size, age groups, and selected joint space metrics. At the same time study findings confirm the clinical significance of joint space; a normal joint space is required for free movement of mandibular condyle along with articular disc. The widening or narrowing of joint space may indicate to some type of TMJ pathology, as well as the difference in the volumes of joint space between the two sides is the cause of facial asymmetry in most cases, while the bony parts of the joints remain symmetrical~\cite{Talmaceanu2018_disorders_tmj}

% MC position lead to differences in measurements
One of the controversial issues in the TMJ study is the determination of ''ideal'' mandibular condyle position. In disputes between gnatologists and orthodontists, the concept of right mandibular condyle position ranged from the most retruded position of the condyle in the glenoid fossa to the most superior position of the condyle~\cite{mc_position}. 
The most common methods for assessment of mandibular condyle positions are based on the assessment of the joint space measurements, between the condyle and the mandibular fossa in special points. A variation of these measurements suggest several condyle positioning systems and leads to differences in the determination of the "gold standard" for these measurements.

A TMJ is characterized by a complex anatomic structure and specific irregularities with the dimensions of the computed tomography (CT) spatial resolution limits, and is among the most complex joints in human body with vast morphological variability \cite{Farias2015_morphology_var}. 
Because of the complexity of TMJ, the use of 2D slice-by-slice visualization is not sufficient, requiring a true 3D reconstruction to describe its anatomy and to find the cause of a given symptom. Yet, many dentists
% , especially in the developing countries, 
have to dismiss the 3D structure and to resort to simple linear measurements of the object dimensions in the 2D images.

% current metrics
Among currently used metrics for TMJ examinations are the horizontal condylar angle (HCA), sagittal ramus angle (SRA), medial joint space (MJS), lateral joint space (LJS), superior joint space (SJS), anterior joint space (AJS), and the width and depth of mandibular fossa (FW, FD)
%~\cite{something}.
Despite the fact that to measure the angles (HCA and SRA) and the linear parameters of joint space (MJS, LJS, SJS, AJS, FW, FD) in 3D case the volumetric reconstruction of bone structures is required, these metrics are only a generalization of the 2D measurements into 3D space, as they are still measurements between two or three points selected by the eye. We suggest paying attention to more volumetrical metrics, such as the volume and surface area of joint space, for the most complete morphologic examination of TMJ.

% Manual vs Auto: compared
% \begin{figure}
%     \centering
%     \includegraphics[width=0.8\textwidth]{EPS/Cavity_compared.eps}
%     \caption{Compared shapes of the manual annotation negative volume (gray) and the inflated shape (green): a) coronal (side view), b) axial (bottom view). Shape differ significantly as the inflated volume doesn't feature unnatural spherical elements found in the manual annotations.}
%     \label{fig:nv_compare}
% \end{figure}

% \bibliographystyle{unsrt}
% % \bibliography{main}

% \end{document}

\bibliographystyle{unsrt}
\bibliography{main}

\begin{thebibliography}{10}

\bibitem{TMJ_mechanics}
Jan~Harm Koolstra.
\newblock Dynamics of the human masticatory system.
\newblock {\em Critical reviews in oral biology and medicine : an official
  publication of the American Association of Oral Biologists}, 13:366--76, 02
  2002.

\bibitem{Wadhwa930_TMJ_disorders}
Sunil Wadhwa and Sunil Kapila.
\newblock Tmj disorders: Future innovations in diagnostics and therapeutics.
\newblock {\em Journal of Dental Education}, 72(8):930--947, 2008.

\bibitem{TMJ_disorder_qualityLife}
Debora de~Melo~Trize, Marcela~Pagani Calabria, Solange de~Oliveira
  Braga~Franzolin, Carolina~Ortigosa Cunha, and Sara~Nader Marta.
\newblock Is quality of life affected by temporomandibular disorders?
\newblock {\em Einstein (S{\~{a}}o Paulo)}, 16(4), 2018.

\bibitem{Farias2015_morphology_var}
J~F~G de~Farias, S~L~S Melo, P~M Bento, L~S A~F Oliveira, P~S~F Campos, and D~P
  de~Melo.
\newblock Correlation between temporomandibular joint morphology and disc
  displacement by {MRI}.
\newblock {\em Dentomaxillofacial Radiology}, 44(7):20150023, sep 2015.

\bibitem{Talmaceanu2018_disorders_tmj}
Daniel Talmaceanu, Lavinia~Manuela Lenghel, Nicolae Bolog, Mihaela Hedesiu,
  Smaranda Buduru, Horatiu Rotar, Mihaela Baciut, and Grigore Baciut.
\newblock Imaging modalities for temporomandibular joint disorders: an update.
\newblock {\em Medicine and Pharmacy Rep.}, 91(3):280--287, 2018.

\bibitem{2d_vs_3d}
Xianchao~Xu Yuanli~Zhang and Zhan Liu.
\newblock Comparison of morphologic parameters of temporomandibular joint for
  asymptomatic subjects using the two-dimensional and three-dimensional
  measuring methods, 2017.

\bibitem{Nove3DTMJ}
Renie Ikeda, Snehlata Oberoi, David~F. Wiley, Christian Woodhouse, Melissa
  Tallman, Wint~Wint Tun, Charles McNeill, Arthur~J. Miller, and David Hatcher.
\newblock Novel 3-dimensional analysis to evaluate temporomandibular joint
  space and shape.
\newblock {\em American Journal of Orthodontics and Dentofacial Orthopedics},
  149(3):416--428, mar 2016.

\bibitem{orthopedics}
Peter Bullough.
\newblock {\em Orthopaedic Pathology}.
\newblock Mosby, 2009.

\bibitem{plastic_surgery}
Eduardo Rodriguez Joseph Losee~Peter Neligan.
\newblock {\em Plastic Surgery: Craniofacial, Head and Neck Surgery and
  Pediatric Plastic Surgery}.
\newblock Elsevier, 2017.

\bibitem{wrist_review}
Stok KS Barnabe C; SPECTRA~Collaboration Nagaraj~S, Finzel~S.
\newblock High-resolution peripheral quantitative computed tomography imaging
  in the assessment of periarticular bone of metacarpophalangeal and wrist
  joints.
\newblock {\em The Journal of Rheumatology}, pages 1921--1934, 2016.

\bibitem{tmj_space}
Aarati~S Panchbhai.
\newblock Temporomandibular joint space.
\newblock {\em Indian Journal of Oral Health and Research}, 2017.

\bibitem{SHEN2017663}
Wei Shen, Mu~Zhou, Feng Yang, Dongdong Yu, Di~Dong, Caiyun Yang, Yali Zang, and
  Jie Tian.
\newblock Multi-crop convolutional neural networks for lung nodule malignancy
  suspiciousness classification.
\newblock {\em Pattern Recognition}, 61:663 -- 673, 2017.

\bibitem{cervical_vertebrae}
Greg~Slabaugh S.~M. Masudur Rahman Al~Arif, Karen~Knapp.
\newblock Region-aware deep localization framework for cervical vertebrae in
  x-ray images.
\newblock In {\em Deep Learning in Medical Image Analysis and Multimodal
  Learning for Clinical Decision Support}, pages 74--82. Springer, Cham, 2017.

\bibitem{roi_aware_unet}
Zi-Xian Wang Li-Zhi Liu Ying Jin Chao-Feng Li Lisheng Wang Hao~Chen
  Yi-Jie~Huang, Qi~Dou and Rui-Hua Xu.
\newblock 3d roi-aware u-net for accurate and efficient colorectal tumor
  segmentation, 2018.

\bibitem{prokopenko2019unpaired}
Denis Prokopenko, Jo{\"e}l~Valentin Stadelmann, Heinrich Schulz, Steffen
  Renisch, and Dmitry~V Dylov.
\newblock Unpaired synthetic image generation in radiology using gans.
\newblock In {\em Workshop on Artificial Intelligence in Radiation Therapy},
  pages 94--101. Springer, Cham, 2019.

\bibitem{skull_segmentation}
Wouter Kouwb Faruk Diblenb-Adriënne Mendrikb Jan~Wolffa Jordi~Minnemaa,
  Maureen van~Eijnattena.
\newblock Ct image segmentation of bone for medical additive manufacturing
  using a convolutional neural network.
\newblock {\em Computers in Biology and Medicine}, pages 130--139, 2018.

\bibitem{multi_task_cascades}
Kaiming~He Jifeng~Dai and Jian Sun.
\newblock Instance-aware semantic segmentation via multi-task network cascades.
\newblock In {\em 2016 IEEE Conference on Computer Vision and Pattern
  Recognition (CVPR)}, pages 3150--3158. {IEEE}, 2016.

\bibitem{mask_rcnn}
Kaiming~He Jifeng~Dai and Jian Sun.
\newblock Mask r-cnn.
\newblock In {\em 2017 IEEE International Conference on Computer Vision
  (ICCV)}, pages 2980--2988. {IEEE}, 2017.

\bibitem{segmentation_by_detection}
Dana Cobzas Martin~Jagersand Min~Tang, Ziehen~Zhang and Jacob~L Jaremko.
\newblock Segmentation-by-detection: A cascade network for volumetric medical
  image segmentation.
\newblock In {\em ISBI 2018}, pages 1356--1359. {IEEE}, 2018.

\bibitem{Lee-DL-overview}
June-Goo Lee, Sanghoon Jun, Young-Won Cho, Hyunna Lee, Guk~Bae Kim, Joon~Beom
  Seo, and Namkug Kim.
\newblock Deep learning in medical imaging: General overview.
\newblock {\em Korean Journal of Radiology}, 18(4):570, 2017.

\bibitem{chowdhury2017blood}
Aritra Chowdhury, Dmitry~V Dylov, Qing Li, Michael MacDonald, Dan~E Meyer,
  Michael Marino, and Alberto Santamaria-Pang.
\newblock Blood vessel characterization using virtual 3d models and
  convolutional neural networks in fluorescence microscopy.
\newblock In {\em IEEE ISBI 2017}, pages 629--632. IEEE, 2017.

\bibitem{Zhou2019}
Tongxue Zhou, Su~Ruan, and St{\'{e}}phane Canu.
\newblock A review: Deep learning for medical image segmentation using
  multi-modality fusion.
\newblock {\em Array}, 3-4:100004, sep 2019.

\bibitem{ronneberger2015unet}
Olaf Ronneberger, Philipp Fischer, and Thomas Brox.
\newblock U-net: Convolutional networks for biomedical image segmentation,
  2015.

\bibitem{iek2016_3dUnet}
Özgün Çiçek, Ahmed Abdulkadir, Soeren~S. Lienkamp, Thomas Brox, and Olaf
  Ronneberger.
\newblock 3d u-net: Learning dense volumetric segmentation from sparse
  annotation, 2016.

\bibitem{milletari2016_Vnet}
Fausto Milletari, Nassir Navab, and Seyed-Ahmad Ahmadi.
\newblock V-net: Fully convolutional neural networks for volumetric medical
  image segmentation.
\newblock In {\em 2016 Fourth International Conference on 3D Vision (3DV)}.
  {IEEE}, oct 2016.

\bibitem{zhou2018unetPlusPlus}
Zongwei Zhou, Md~Mahfuzur~Rahman Siddiquee, Nima Tajbakhsh, and Jianming Liang.
\newblock Unet++: A nested u-net architecture for medical image segmentation,
  2018.

\bibitem{zhang2018mdunet}
Jiawei Zhang, Yuzhen Jin, Jilan Xu, Xiaowei Xu, and Yanchun Zhang.
\newblock Mdu-net: Multi-scale densely connected u-net for biomedical image
  segmentation, 2018.

\bibitem{Sevastopolsky2019}
Artem Sevastopolsky, Stepan Drapak, Konstantin Kiselev, Blake~M. Snyder,
  Jeremy~D. Keenan, and Anastasia Georgievskaya.
\newblock Stack-u-net: refinement network for improved optic disc and cup image
  segmentation.
\newblock In {\em Medical Imaging 2019: Image Processing}. {SPIE}, 2019.

\bibitem{oktay2018_attentionUnet}
Ozan Oktay, Jo~Schlemper, Loic~Le Folgoc, Matthew Lee, Mattias Heinrich,
  Kazunari Misawa, Kensaku Mori, Steven McDonagh, Nils~Y Hammerla, Bernhard
  Kainz, Ben Glocker, and Daniel Rueckert.
\newblock Attention u-net: Learning where to look for the pancreas, 2018.

\bibitem{MulticlassWTMJ}
Yunhe Gao, Rui Huang, Ming Chen, Zhe Wang, Jincheng Deng, Yuanyuan Chen, Yiwei
  Yang, Jie Zhang, Chanjuan Tao, and Hongsheng Li.
\newblock {FocusNet}: Imbalanced large and small organ segmentation with an
  end-to-end deep neural network for head and neck {CT} images.
\newblock In {\em Lecture Notes in Computer Science}, pages 829--838. Springer
  International Publishing, 2019.

\bibitem{ACCVinflation}
Andrei Zaharescu, Edmond Boyer, and Radu Horaud.
\newblock {TransforMesh} : A topology-adaptive mesh-based approach to surface
  evolution.
\newblock In {\em {ACCV} 2007}, pages 166--175. Springer, 2007.

\bibitem{CountourSegment2018}
R.J. Hemalatha, T.R. Thamizhvani, A.~Dhivya, Josline Joseph, Bincy Babu, and
  R.~Chandrasekaran.
\newblock Active contour based segmentation techniques for medical image
  analysis.
\newblock {\em Medical and Biological Image Analysis}, 07 2018.

\bibitem{NestedMesh2017}
Alec Jacobson.
\newblock Generalized matryoshka: Computational design of nesting objects.
\newblock {\em Computer Graphics Forum}, 36(5):27--35, 2017.

\bibitem{TMJ_FW_lateral}
Rodrigo~Lorenzi Poluha, Carolina~Ortigosa Cunha, Leonardo~Rigoldi Bonjardim,
  and Paulo César~Rodrigues Conti.
\newblock Temporomandibular joint morphology does not influence the presence of
  arthralgia in patients with disk displacement with reduction: a magnetic
  resonance imaging–based study.
\newblock {\em Oral Surgery}, 129(2):149 -- 157, 2020.

\bibitem{Skouras2014_Inflation}
M{\'{e}}lina Skouras, Bernhard Thomaszewski, Peter Kaufmann, Akash Garg, Bernd
  Bickel, Eitan Grinspun, and Markus Gross.
\newblock Designing inflatable structures.
\newblock {\em {ACM} Transactions on Graphics}, 33(4):1--10, jul 2014.

\bibitem{VNET_MICCAI18_WORKSHOP}
Gustavo~Retuci Pinheiro, Raphael Voltoline, Mariana~P. Bento, and
  Let{\'{\i}}cia Rittner.
\newblock V-net and u-net for ischemic stroke lesion segmentation in a small
  dataset of perfusion data.
\newblock In {\em BrainLes 2018, Granada, Spain, September 16, 2018, Part {I}},
  volume 11383 of {\em Lecture Notes in Computer Science}, pages 301--309.
  Springer, 2018.

\bibitem{DLReview_MED}
Mohammad~Hesam Hesamian, Wenjing Jia, Xiangjian He, and Paul Kennedy.
\newblock Deep learning techniques for medical image segmentation: Achievements
  and challenges.
\newblock {\em Journal of Digital Imaging}, 32(4):582--596, may 2019.

\bibitem{wasserstein_dist}
C{\'{e}}dric Villani.
\newblock The wasserstein distances.
\newblock In {\em Grundlehren der mathematischen Wissenschaften}, pages
  93--111. Springer, 2009.

\bibitem{knee_segmentation}
Q.~H.~Zhang Y.~Sun, E. C.~Teo.
\newblock Discussions of knee joint segmentation.
\newblock In {\em 2006 International Conference on Biomedical and
  Pharmaceutical Engineering}. {IEEE}, 2006.

\bibitem{knee_box}
Noel E.~O'Connor Joseph~Antony, Kevin~McGuinness.
\newblock Automatic detection of knee joints and quantification of knee
  osteoarthritis severity using convolutional neural networks.
\newblock {\em Machine Learning and Data Mining in Pattern Recognition}, 2017.

\bibitem{jsw1}
Michelle Kan Eva Szabo Susan G. Barr Liam Martin Steven K.~Boyd Cheryl~Barnabe,
  Helen~Buie.
\newblock Reproducible metacarpal joint space width measurements using 3d
  analysis of images acquired with high-resolution peripheral quantitative
  computed tomography.
\newblock {\em Med Eng Phys.}, 35(10):1540–1544, 2013.

\bibitem{jsw2}
Youssof E Locrelle H Thomas T Chapurlat R Marotte~H. Boutroy~S, Hirschenhahn~E.
\newblock Importance of hand positioning in 3d joint space morphology
  assessment.
\newblock {\em Arthr Rheum}, 2013.

\bibitem{jsw3}
Lee C.H. Kuo D. et~al. Burghardt, A.J.
\newblock Quantitative in vivo hr-pqct imaging of 3d wrist and
  metacarpophalangeal joint space width in rheumatoid arthritis.
\newblock {\em Ann Biomed Eng}, 41:2553–2564, 2013.

\bibitem{metacarpophalangeal}
Stephanie Boutroy Michiel P. H. Peters Sarah L. Manske Vincent Stadelmann
  Nicolas Vilayphiou Joop P. van den Bergh Piet Geusens Xiaojuan Li Hubert
  Marotte Bert van Rietbergen Steven K. Boyd Cheryl Barnabe; for the
  SPECTRA~Collaboration Kathryn S.~Stok, Andrew J.~Burghardt.
\newblock Consensus approach for 3d joint space width of metacarpophalangeal
  joints of rheumatoid arthritis patients using high-resolution peripheral
  quantitative computed tomography.
\newblock {\em Quantitative Imaging in Medicine and Surgery}, 10(2), 2020.

\bibitem{mc_position}
Carlos A.Pires Maria J. Ponces Jorge D.~Lopes Eugénio~Martins, Joana C.~Silva.
\newblock Sagittal joint spaces of the temporomandibular joint: Systematic
  review and meta-analysis.
\newblock {\em Journal of Central South University (Medical Sciences)}, 2015.

\end{thebibliography}

% \section*{References}

% References follow the acknowledgments. Use unnumbered first-level heading for
% the references. Any choice of citation style is acceptable as long as you are
% consistent. It is permissible to reduce the font size to \verb+small+ (9 point)
% when listing the references.
% {\bf Note that the Reference section does not count towards the eight pages of content that are allowed.}
% \medskip

% \small

% [1] Alexander, J.A.\ \& Mozer, M.C.\ (1995) Template-based algorithms for
% connectionist rule extraction. In G.\ Tesauro, D.S.\ Touretzky and T.K.\ Leen
% (eds.), {\it Advances in Neural Information Processing Systems 7},
% pp.\ 609--616. Cambridge, MA: MIT Press.

% [2] Bower, J.M.\ \& Beeman, D.\ (1995) {\it The Book of GENESIS: Exploring
%   Realistic Neural Models with the GEneral NEural SImulation System.}  New York:
% TELOS/Springer--Verlag.

% [3] Hasselmo, M.E., Schnell, E.\ \& Barkai, E.\ (1995) Dynamics of learning and
% recall at excitatory recurrent synapses and cholinergic modulation in rat
% hippocampal region CA3. {\it Journal of Neuroscience} {\bf 15}(7):5249-5262.
% \include{supplement}
\end{document}